\documentclass[%
 aip,
 amsmath,amssymb,
 reprint,%
]{revtex4-1}

\usepackage{graphicx}% Include figure files
\usepackage{dcolumn}% Align table columns on decimal point
\usepackage{bm}% bold math
\usepackage{enumitem}

\usepackage[utf8]{inputenc}
\usepackage[T1]{fontenc}
\usepackage{mathptmx}
\usepackage{etoolbox}

\makeatletter
\def\@email#1#2{%
 \endgroup
 \patchcmd{\titleblock@produce}
  {\frontmatter@RRAPformat}
  {\frontmatter@RRAPformat{\produce@RRAP{*#1\href{mailto:#2}{#2}}}\frontmatter@RRAPformat}
  {}{}
}%
\makeatother
\begin{document}

\preprint{AIP/123-QED}

\title[On Hyperbolic Attractors in Complex Shimizu -- Morioka Model]{On Hyperbolic Attractors in Complex Shimizu -- Morioka Model}
% Force line breaks with \\
\author{V.P. Kruglov}
 %%\altaffiliation[Also at ]{Institute of Electronic Engineering and Instrumentation, Yuri Gagarin State Technical University of Saratov, Politekhnicheskaya 77, Saratov 410054, Russia}%Lines break automatically or can be forced with \\
 \email{kruglovyacheslav@gmail.com}

\author{I.R. Sataev}%
 \email{sataevir@gmail.com}
\affiliation{Kotelnikov’s Institute of Radio-Engineering and Electronics of RAS, Saratov Branch, Zelenaya 38, Saratov, 410019, Russia%\\This line break forced with \textbackslash\textbackslash
}%

\date{\today}% It is always \today, today,

\begin{abstract}
We present a modified complex-valued Shimizu -- Morioka system with uniformly hyperbolic attractor.
The numerically observed attractor in Poincar\'{e} cross-section is topologically close to Smale -- Williams solenoid.
The arguments of the complex variables undergo Bernoulli-type map, essential for Smale -- Williams attractor, due to the geometrical arrangement of the phase space and an additional perturbation term.
The transformation of the phase space near the saddle equilibrium ``scatters'' trajectories to new angles, then trajectories run from the saddle and return to it for the next ``scatter''.
We provide the results of numerical simulations of the model and demonstrate typical features of the appearing hyperbolic attractor of Smale -- Williams type.
Importantly, we show in numerical tests the transversality of tangent subspaces -- a pivotal property of uniformly hyperbolic attractor.
\end{abstract}

\maketitle

\begin{quotation}
Lorenz attractor is the brand name of dynamical chaos. 
It appears in different physical problems, from the models of atmosphere instability to lasers and mechanical systems. 
The simplest equations with Lorenz attractors are Shimizu -- Morioka system. 
It is three-dimensional and relatively simple. 
Unfortunately, from the mathematical point of view it is not the best kind of chaotic attractor. 
Even before the breakthrough of Edward Lorenz mathematicians Smale, Anosov and Plykin suggested examples of uniformly hyperbolic attractors. 
In this article we show the modification of the classical Shimizu -- Morioka system, that contains uniformly hyperbolic attractor of Smale -- Williams type.
This is actually a development of our previous work: earlier we introduced simple 4-dimensional autonomous flow model with complex variables. 
We constructed it from scratch and explained the phase space transformations that lead to Smale -- Williams attractor formation in Poincar\'{e} cross-section. 
We coined the most important transformation ``the scattering'' of trajectories on the saddle equilibrium. 
The only flaw is that the model has no physical meaning obvious to us. 
We realised that we could construct much more significant complex Shimizu -- Morioka system with hyperbolic attractor using the same approach. 
\end{quotation}

\section{\label{sec1:level1}introduction}

Uniformly hyperbolic attractors are the most refined geometrical shapes of chaos
~\cite{anosov2006dynamical,smale1967differentiable,anosov1995dynamical,katok1997introduction,shilnikov1997mathematical,kuznetsov2011dynamical,kuznetsov2012hyperbolic} .
They are rigorously proved to be genuine chaotic attractors.
All of the trajectories that belong to uniformly hyperbolic attractor are of saddle type with the same dimensions of stable, neutral and unstable subspaces.
Most importantly, the tangent subspaces are transversal to each other at every point on the uniformly hyperbolic attractor and in its vicinity.
These fine geometrical features ensure that the attractor is structurally stable: it preserves its structure under small and even finite perturbations of the governing equations.
It is worth to mention, that there is another wider class of genuine chaotic attractors called pseudohyperbolic
~\cite{turaev1998example,gonchenko2018elements}.
They are not structurally stable, but still preserve their most important features under perturbations.
The definition of pseudohyperbolicity is less strict, while the uniformly hyperbolic attractors are their special subclass.
The famous Lorenz attractor~\cite{lorenz1963deterministic} is pseudohyperbolic~\cite{turaev2008pseudohyperbolicity}, though it is not uniformly hyperbolic, but singular: it contains the saddle equilibrium with its unstable separatrices.
%%These separatrices go through the body of the attractor, therefore they can not be excluded from it. 
There is an open set of parameter values, at which the separatrices become bi-asymptotic to the saddle. 
Due to this fact the Lorenz attractor is not structurally stable, even though it remains intact under changes of parameters. 
There are also wild pseudohyperbolic attractors~\cite{turaev1998example,gonchenko2021wild}: they allow tangencies between subspaces, unlike uniformly hyperbolic attractors, 
but do not generate stable orbits under perturbations. 
Pseudohyperbolic attractors are robust as well as uniformly hyperbolic. 
In comparison, the so called quasi-attractors~\cite{Afraimovich1983strange} are not genuine and are not robust (while they are the most common in applications), because they manifest zero angles between subspaces and birth of stable orbits at any small perturbation.

\begin{figure}[!t]
\includegraphics[width=.49\textwidth,keepaspectratio]{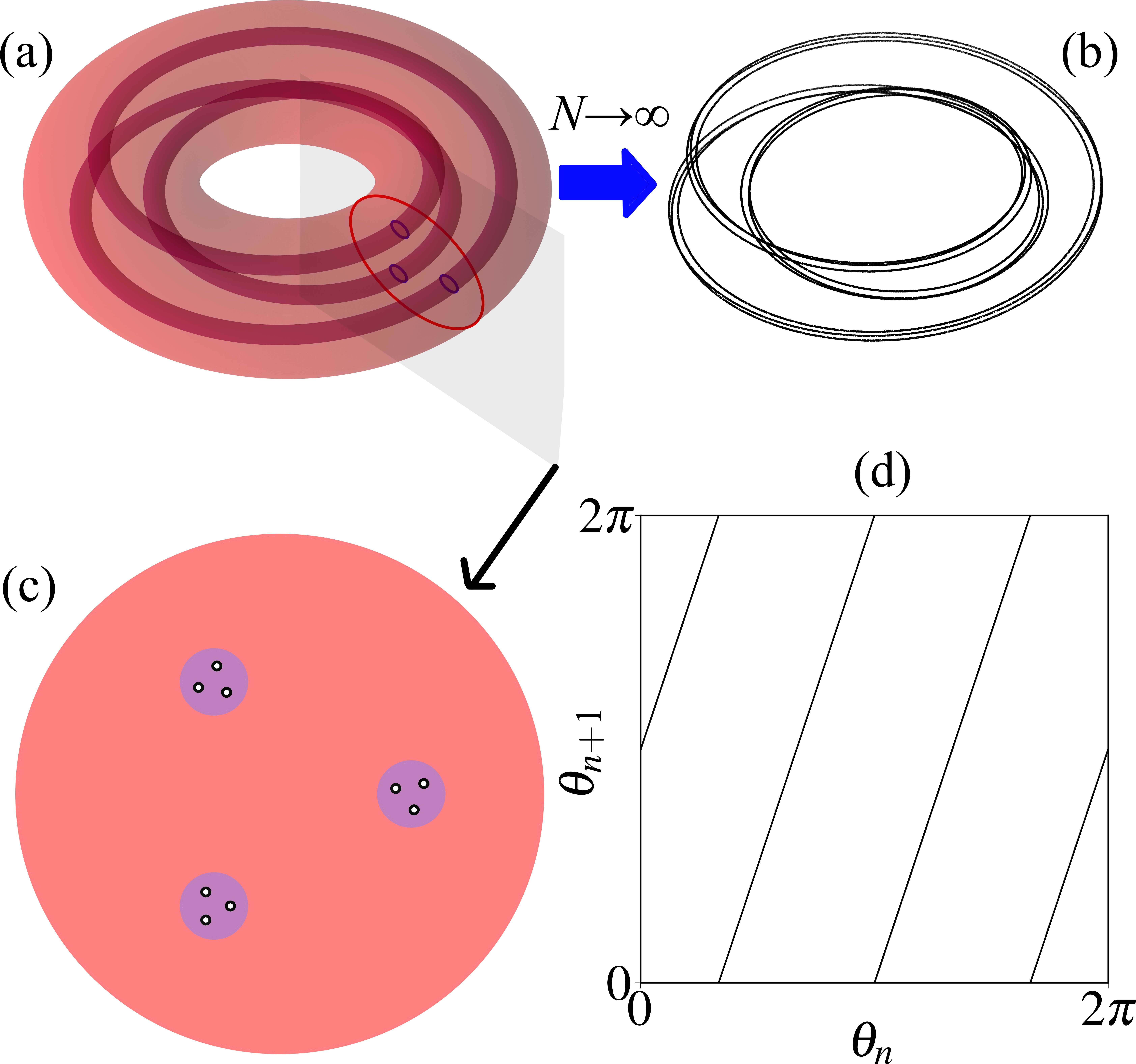}
\caption{
\footnotesize
(a) The action of the map stretches the toroidal domain $3$ times in a direction around the hole, but contracts strongly in other directions and folds inside the initial domain; 
(b) the solenoid with expansion factor $3$ appears in the limit $n\rightarrow\infty$; 
(c) the images of the absorbing domain are shown in transversal cut, in the limit the Cantor set appears; 
(d) the $\theta_n$ vs. $\theta_{n+1}$ diagram for Bernoulli map $\theta_{n+1}=3\theta_n+\pi\,\pmod{2\pi}$, corresponding to previous pictures.
$\pi$ shift is also observed in our complex Shimizu -- Morioka model.
The exact equations to produce these pictures can be found for example in our previous article~\cite{kuznetsov2021smale} on the topic.
}
\label{fig01}
\end{figure}

Smale -- Williams solenoid is one of the conceptual geometrical examples of uniformly hyperbolic attractors
~\cite{smale1967differentiable,williams1974expanding}.
It appears in phase spaces of dimension 3 (or more) under action of a diffeomorphism expanding the space integer times ($2$, $3$, etc.) in angular direction, contracting the space in all transversal dimensions and folding inside.
The result of the infinite iterations is an attractor consisting of infinite number of loops ordered with transversal structure of the Cantor set.
Actually, there is an angular variable $\theta$ undergoing the Bernoulli map.
Fig.~\ref{fig01} shows the action of the map with expansion factor $3$ (panel a), the resulting attractor (panel b), the transversal Cantor set structure of its filaments (panel c) and the $\theta_{n+1}$ vs. $\theta_n$ diagram for the kind of Bernoulli map $\theta_{n+1}=3\theta_n+\pi\,\pmod{2\pi}$.
Attractors similar to Smale -- Williams solenoid appear in Poincar\'{e} cross-sections of specially designed by Kuznetsov mathematical and physical models (electronic or mechanical)
~\cite{kuznetsov2005example,kuznetsov2007autonomous,kuznetsov2020mechanical,kuznetsov2012hyperbolic}
, where phase of the oscillations is usually the angular variable under Bernoulli map.
The hyperbolicity of the attractor of model from~\cite{kuznetsov2005example} have been confirmed by computer assisted proofs
~\cite{kuznetsov2007hyperbolic,wilczak2010uniformly}.

Recently we have studied a peculiar system of differential equations with two complex variables, where the emergence of hyperbolic attractor of Smale -- Williams type is observed in numerics and is geometrically interpretable and explainable~\cite{kuznetsov2021smale}.
The model is derived from the 2D self-oscillatory real-valued system with homoclinic bifurcation of the saddle equilibrium: at some parameter value stable limit cycles glue into the separatrix loops forming the figure-8 contour~\cite{shilnikov1968generation}.
In the complexified system there is a saddle equilibrium at the origin with two equal positive eigenvalues and two equal negative eigenvalues.
There is also a special perturbation providing integer times expansion of the arguments of complex variables (on $2\pi$ average), when the trajectory passes near the saddle.
In the vicinity of the saddle the trajectory turns to an angle governed by map close to Bernoulli map.
We have checked numerically that the stable and unstable tangent subspaces of attractor in Poincar\'{e} cross-section are always transversal, also we have observed the structural stability and other properties of uniformly hyperbolic attractor.
Therefore we have concluded that the system indeed possesses the hyperbolic attractor of Smale -- Williams type.
At the same time we do not know the physical interpretation of this clearly mathematical example.

There is a well-known example of homoclinic bifurcation in systems of equations originated in physics.
It is the homoclinic butterfly bifurcation in Lorenz and Shimizu -- Morioka systems~\cite{shil1993bifurcations,shil1993normal}.
Both of the systems manifest genuine Lorenz attractor, forming from unstable set that arises after the saddle limit cycles emerge from the separatrix loops of the saddle~\cite{afraimovich1977origin,williams1979structure,guckenheimer1979structural}.
But there is also a bifurcation of stable limit cycles emerging from the separatrix loops, very similar to our previous example~\cite{kuznetsov2021smale}.
We start with Shimizu -- Morioka system since it is better suited for our goals.
It is 3D and mathematically of the form of nonlinear oscillator with an additional relaxation variable.
We change the two of the three variables from real to complex and supply an additional perturbation term.
Remarkably our modified 5D Shimizu -- Morioka system manifests uniformly hyperbolic attractor of Smale -- Williams type, while the classical 3D system manifests the pseudohyperbolic Lorenz attractor. 
In our research we have learned surprisingly that the new 5D system does not have any pseudohyperbolic attractors at all, though a very Lorenz-looking-like attractor still exists. 

Complex-valued Lorenz and Shimizu -- Morioka systems have been studied before~\cite{vladimirov1998properties,vladimirov1998complex}, 
they appear naturally in physical problems: of detuned lasers~\cite{ning1990detuned}, of baroclinic instability in two-layered rotating fluid~\cite{gibbon1982real,fowler1982complex,gibbon1980derivation}. 
It is stated in~\cite{fowler1983real}, that real and complex Lorenz models appear naturally in dispersive unstable systems with weak dissipation, in contrast with truncated model of two-dimensional convection with high dissipation, studied by Lorenz. 
Complex Lorenz and Shimizu -- Morioka systems are symmetric under transformation $\left(x,\,y,\,z\right)\rightarrow\left(xe^{i\phi},\,ye^{i\phi},\,z\right)$.
Our modification by adding non-small perturbation removes this symmetry. 
%%At the time we have no physical motivation to add this new term, but we notice, that it belongs to the class of holomorphic functions. 

In Section II we discuss the classical Shimizu -- Morioka system and construct its complex modification. 
In Section III we provide the results of numerical studies of the complex Shimizu -- Morioka system: portraits of attractors and Lyapunov exponents. 
In Section IV we discuss the criteria of angles -- the technique to determine, if the attractor is uniformly hyperbolic, pseudohyperbolic or a quasi-attractor. 
We also give the results of the test applied to our model. 
The charts of regimes are also provided. 

\section{\label{sec2:level1} The composition of complex-valued Shimizu -- Morioka equations}

\begin{figure}[!t]
\includegraphics[width=.49\textwidth,keepaspectratio]{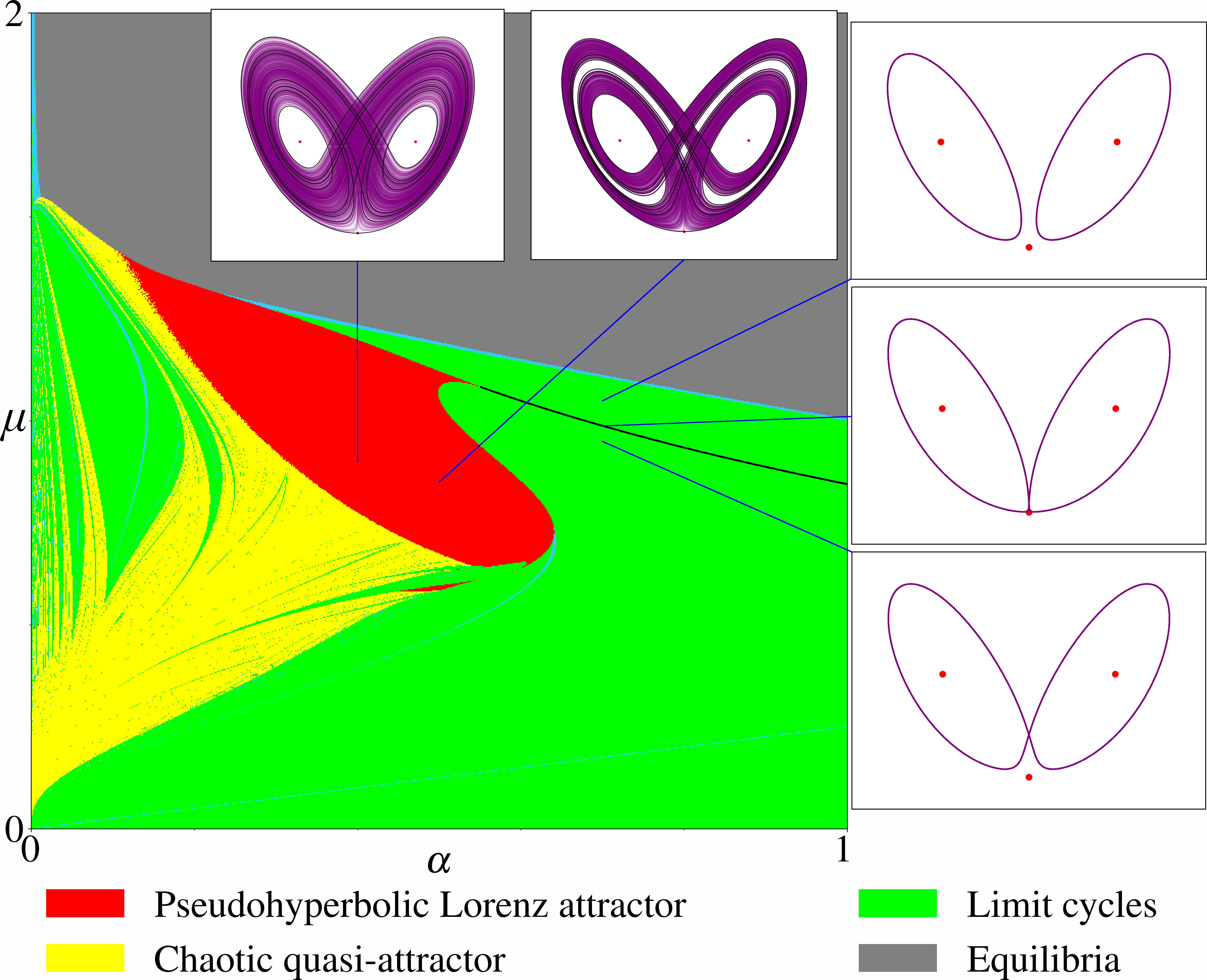}
\caption{
\footnotesize
The chart of regimes of Shimizu -- Morioka model~\eqref{eq01}, $\mu$ vs. $\alpha$, obtained numerically by calculating of Lyapunov exponents and by checking of the transversality of tangent subspaces.
In the yellow regions chaotic quasi-attractors appear. %%: the largest Lyapunov exponent is positive for typical trajectories, but the angles between tangent subspaces are zero at some points on attractor (see section ...).
In the red region the pseudohyperbolic Lorenz attractors manifest. %% appear: the angles between tangent subspaces are never zero in our numerical tests.
%%One can see that the region of Lorenz attractor appears continuous, thus the attractor is robust, though it is not structurally stable. 
In the green regions there is no chaos, but stable limit cycles appear. 
In the grey region the only attractors are equilibria.
Two upper panels contain the examples of pseudohyperbolic Lorenz attractors: at $\mu=0.9$, $\alpha=0.4$ and at $\mu=0.85$, $\alpha=0.5$.
Three side panels demonstrate the stable homoclinic butterfly at $\mu\approx 0.98695\ldots$, $\alpha=0.7$ (the saddle at the origin has negative saddle value) and two phase space configurations near it.
The black line outlines the homoclinic bifurcation with negative saddle value.
See articles~\cite{shil1993bifurcations,shil1993normal,barrio2012kneadings,xing2014symbolic} to compare the charts.
}
\label{fig02}
\end{figure}

The real-valued Shimizu -- Morioka system~\cite{shimizu1980bifurcation} is
\begin{equation}
\begin{aligned}
\dot{x}&=y,\\
\dot{y}&=-\mu y+\left(1-z\right) x,\\
\dot{z}&=-\alpha z+x^2.
\end{aligned}
\label{eq01}
\end{equation}
The first two equations compose a nonlinear oscillator, while the third equation governs a feedback loop through the additional variable $z$.
The Shimizu -- Morioka system is very similar to Lorenz model. 
Indeed, one can transform the Lorenz equations to Shimizu -- Morioka~\cite{shil1993normal} with an additional cubic term $x^3$ in the equation for $y$.

The Shimizu -- Morioka system has been thoroughly investigated, see~\cite{shil1993bifurcations,tigan2011analytical,capinski2018computer} for example.
The system~\eqref{eq01} is invariant with respect to $\left(x,\,y,\,z\right)\rightarrow\left(-x,\,-y,\,z\right)$.
It possesses three equilibria in Shimizu -- Morioka system for $\mu>0,\,\alpha>0$: the saddle $O=\left(0,\,0,\,0\right)$ and
two foci or saddle-foci $O_{1,2}=\left(\pm\sqrt{\alpha},\,0,\,1\right)$.

It is proved, that Shimizu -- Morioka system possesses genuine Lorenz attractor~\cite{capinski2018computer}.
The scenario of the formation of Lorenz attractor involves the homoclinic bifurcation of the saddle equilibrium with positive saddle value 
$\sigma=\lambda_1+\lambda_2$ (where $\lambda_1$ is the positive eigenvalue of the saddle and $\lambda_2$ is closest to zero negative eigenvalue).
We on the other hand are interested in homoclinic bifurcation with negative saddle value.
It is accompanied by gluing of stable limit cycles into bi-asymptotic separatrices of the saddle.
In classical three-dimensional Shimizu -- Morioka system such bifurcation does not lead to formation of chaotic attractor, but it does in our modification.

Fig.~\ref{fig02} shows the chart of regimes of system~\eqref{eq01}. 
The thick black curve outlines the stable homoclinic bifurcation (with negative saddle value). 
The inserted panels on the side demonstrate the homoclinic butterfly at $\alpha=0.7$, $\mu\approx 0.98695\ldots$ and the phase space configurations near it.
We have evaluated numerically the bifurcation line by crude algorithm: the parameter space has been scanned for situations where limit cycles start to visit both positive and negative parts of the phase space relative to plane $x=0$.
Other regimes have been checked by calculation of the Lyapunov exponents with standart methods~\cite{benettin1980lyapunov,shimada1979numerical,pikovsky2016lyapunov}. 
If the largest Lyapunov exponent is negative, the attractor is simple equilibrium, one of two foci $O_{1,2}$ (marked grey on the chart). 
If the largest Lyapunov exponent is zero up to numerical accuracy, the attractor is a limit cycle (marked green).
If the largest Lyapunov exponent is positive, the attractor is chaotic.
We have performed a special test for pseudohyperbolicity, developed in~\cite{kuptsov2012fast,kuptsov2018lyapunov}, we discuss it in detail in the Section IV.
Pseudohyperbolic Lorenz attractors are marked red and chaotic quasi-attractors are marked yellow on the chart.
One can find similar charts in~\cite{shil1993bifurcations,shil1993normal,barrio2012kneadings,xing2014symbolic}.
The region of genuine Lorenz attractors is continuous, because the existence of a Lorenz attractor is a robust property~\cite{afraimovich1983attracting}.
However in this paper we are interested in construction of uniformly hyperbolic attractor, and the test for pseudohyperbolicity is only carried out simultaneously with the test for uniform hyperbolicity.

We change real variables $x$ and $y$ to complex and add a perturbation $\varepsilon y^3$, similar to equations in~\cite{kuznetsov2021smale}:
\begin{equation}
\begin{aligned}
\dot{x}&=y,\\
\dot{y}&=-\mu y+\left(1-z\right) x+\varepsilon y^3,\\
\dot{z}&=-\alpha z+|x|^2.
\end{aligned}
\label{eq02}
\end{equation}
Perturbations $\varepsilon y^2$ and $\varepsilon y^4$ are also possible, but they do not preserve the symmetry $\left(x,\,y,\,z\right)\rightarrow\left(-x,\,-y,\,z\right)$,
while $\varepsilon y^3$ does. 
We preserve this symmetry to obtain more homogeneous distribution of the natural measure on the attractor.
Only the absolute value $|x|^2$ enters the equation for $z$, since $z$ is real variable.
The system~\eqref{eq02} has five real variables overall. 
We restrict the parameters $\mu$, $\alpha$ and $\varepsilon$ to real values. 
The saddle at the origin has two pairs of equal eigenvalues:
$\lambda_{1,2}=-\frac{\mu}{2}-\frac{\sqrt{\mu^2+4}}{2}<0$ and 
$\lambda_{3,4}=-\frac{\mu}{2}+\frac{\sqrt{\mu^2+4}}{2}>0$; there is also eigenvalue $\lambda_5=-\alpha<0$.

Suppose the parameters are close to the values for stable homoclinic butterfly in the original system.
%%Let us track some trajectory approaching the vicinity of the saddle equilibrium at the origin.
While the original system~\eqref{eq01} supports only stable limit cycles at such parameters, 
our modified complex system~\eqref{eq02} manifests instability along the argument of complex variable $z$. 
The only term in~\eqref{eq02}, that changes the arguments of variables, is the perturbation $\varepsilon y^3$.

Let us discuss qualitatively the behavior of the trajectory falling to the vicinity of the saddle. 
the tangent space of the saddle decomposes into two complex planes, spanned by eigenvectors $\textbf{e}_{1,2}$ and $\textbf{e}_{3,4}$ corresponding to $\lambda_{1,2}$ and $\lambda_{3,4}$, and into the direction of exponential compression with $\lambda_5$. 
We consider the variable $z$ unimportant in our following argumentations. 
The direction of the trajectory falling to the saddle is described by some ``incident'' angle $\theta_i$: 
$\textbf{e}_1\cos\theta_i+\textbf{e}_2\sin\theta_i$.
The argument of complex variables $x$ is very close to $\theta_i$: $x\propto e^{i\theta_i}$, the argument of $y$ differs by $\pi$: 
$y=\dot{x}\propto i\dot{\theta_i} e^{i\theta_i}\propto e^{i\left(\theta_i+\pi\right)}$.
After the passage near the saddle the argument of $x$ changes to $\theta_s\approx 3\theta_i+\pi$ --- the ``scattering'' angle.
We call this phenomenon the ``scattering'' of trajectories on the saddle~\cite{kuznetsov2021smale}.
Far from the saddle the arguments of variables are almost constant, because nonlinear terms dominate, other then $\varepsilon y^3$.
The trajectories run similar to original model~\eqref{eq01} when far from the saddle. 
Therefore after every return to the vicinity of the saddle the arguments of variables undergo a transformation very close to the Bernoulli map:
$\theta_{n+1}=3\theta_n+\pi\,\pmod{2\pi}$.
The other directions in the phase space are stable except the neutral direction along the trajectory.
We conclude, that the emergence of attractor similar to Smale -- Williams solenoid is possible in Poincar\'{e} cross-section of complex system~\eqref{eq02} with adequate parameters.

In our reasoning the perturbation $\varepsilon y^3$ is not small. 
However it is possible to apply a perturbative approach with small $\varepsilon$ to phenomenological simplified equations, that grasp the essence of dynamics near the saddle:
\begin{equation}
\begin{aligned}
\dot{x}&=y,\\
\dot{y}&=-\mu y+x+\varepsilon y^3,\\
\dot{z}&=-\alpha z,
\end{aligned}
\label{eq03}
\end{equation}
some nonlinear terms from~\eqref{eq02} are considered negligibly small near the saddle, but the perturbation $\varepsilon y^3$ is retained due to its importance.
The analysis of exactly the same equations has been performed in~\cite{kuznetsov2021smale} (one can omit the equation $\dot{z}=-\alpha z$).
We do not reproduce the argumentations here.

In the next Section we provide the results of numerical investigations of the system~\eqref{eq02}.

\section{\label{sec3:level1} Results of numerical simulations of complex-valued Shimizu -- Morioka equations}

\begin{figure}[!t]
\centering
\includegraphics[width=.23\textwidth,keepaspectratio]{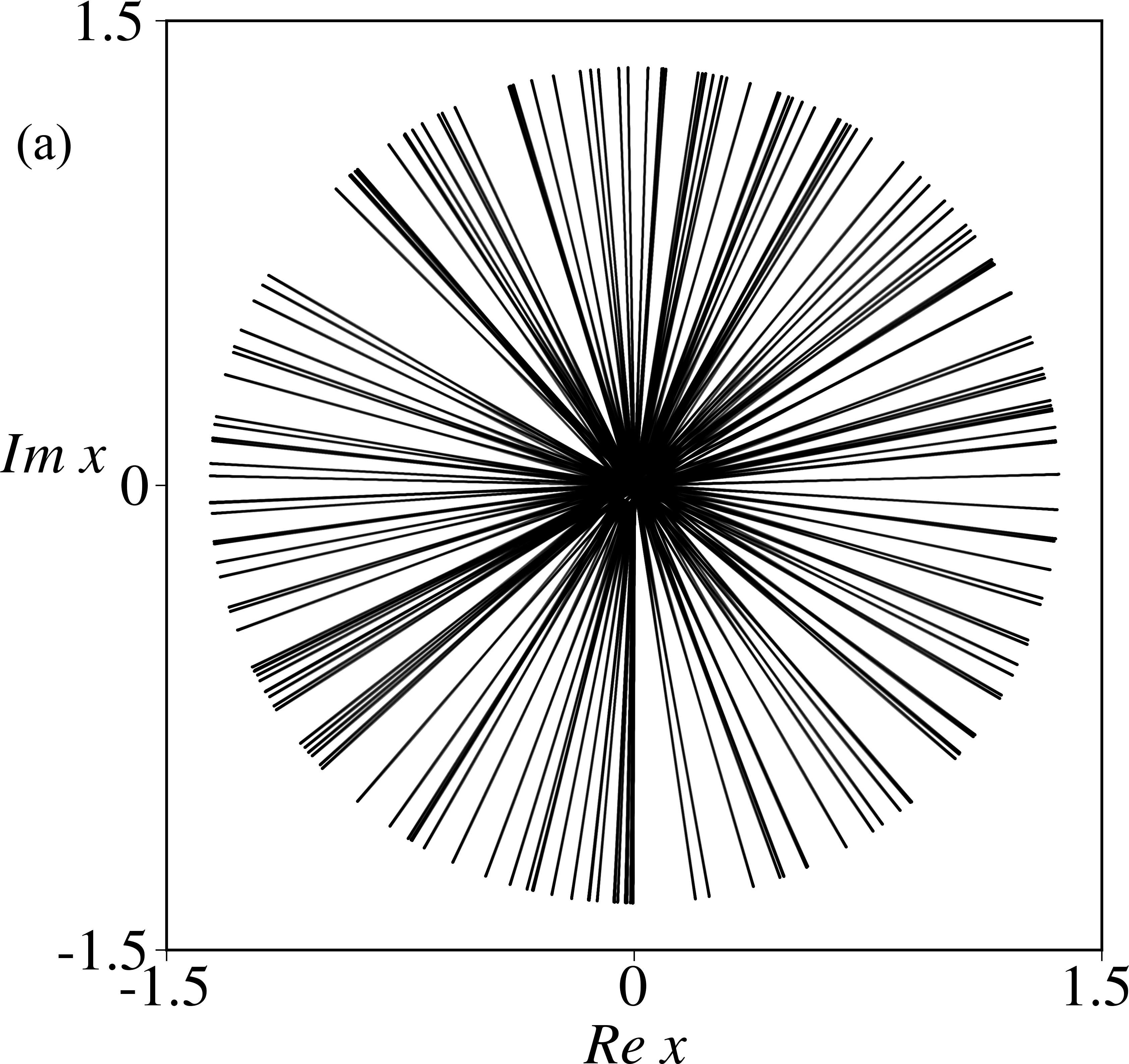}
\includegraphics[width=.23\textwidth,keepaspectratio]{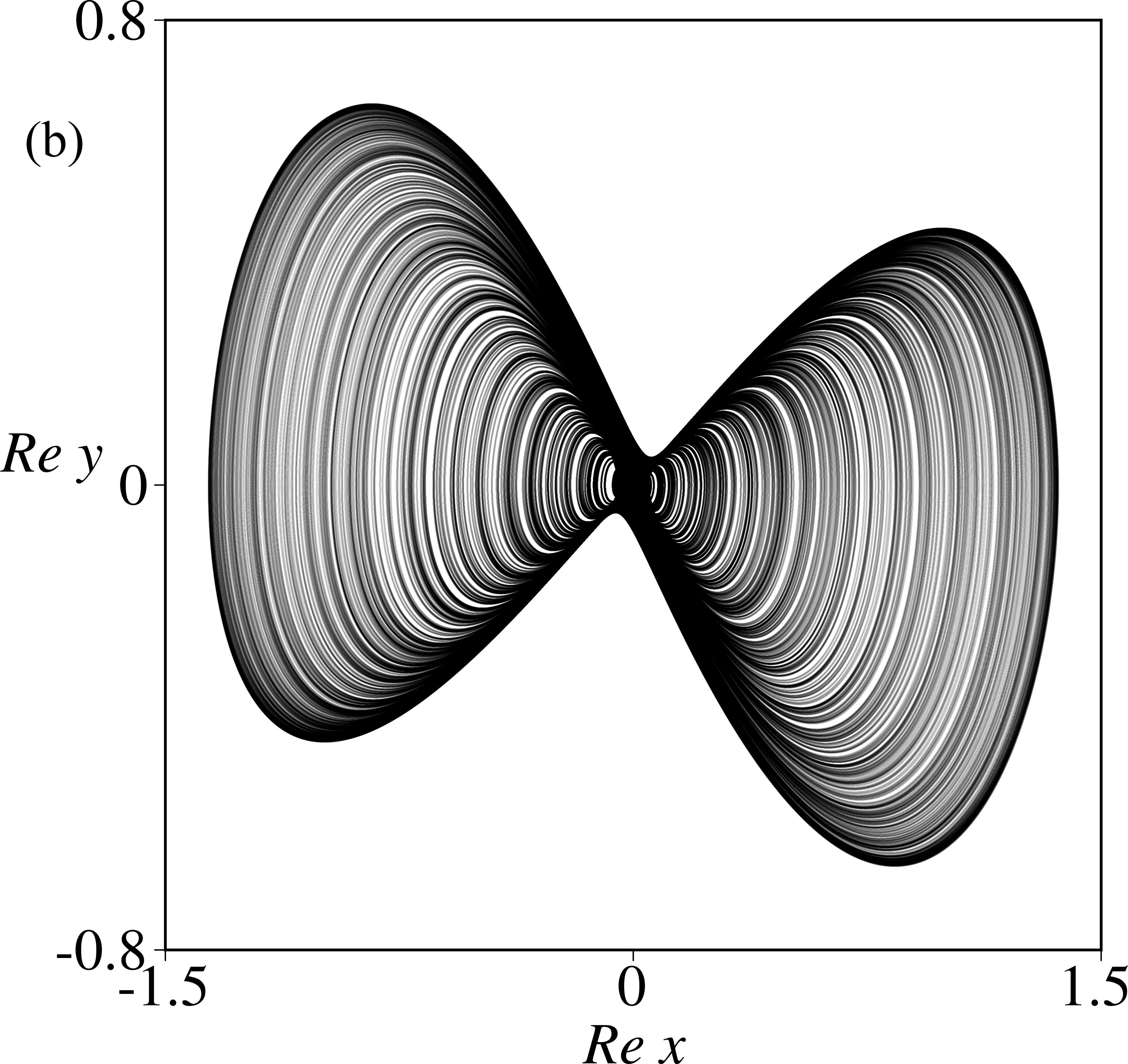}
\includegraphics[width=.215\textwidth,keepaspectratio]{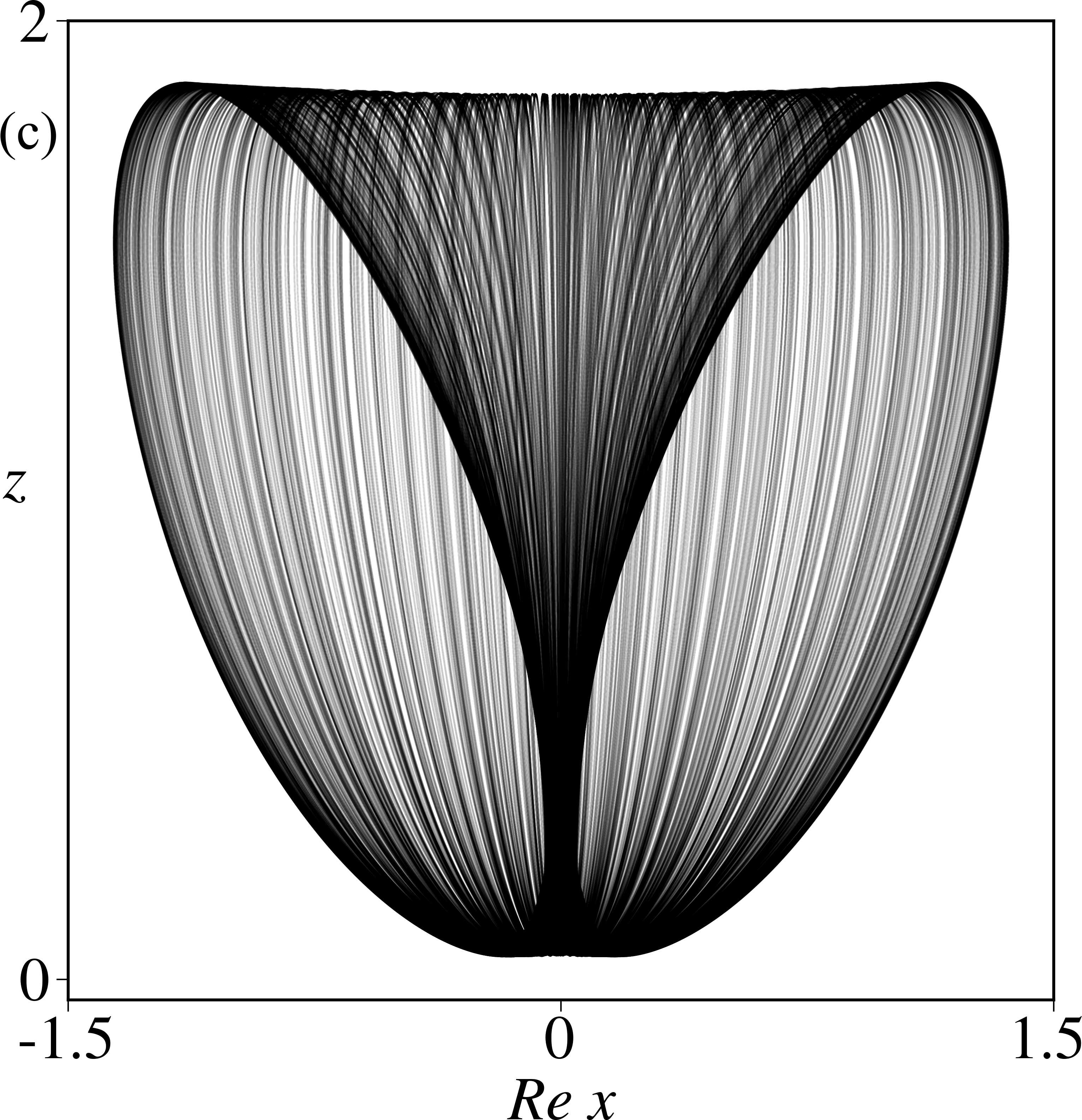}
\includegraphics[width=.23\textwidth,keepaspectratio]{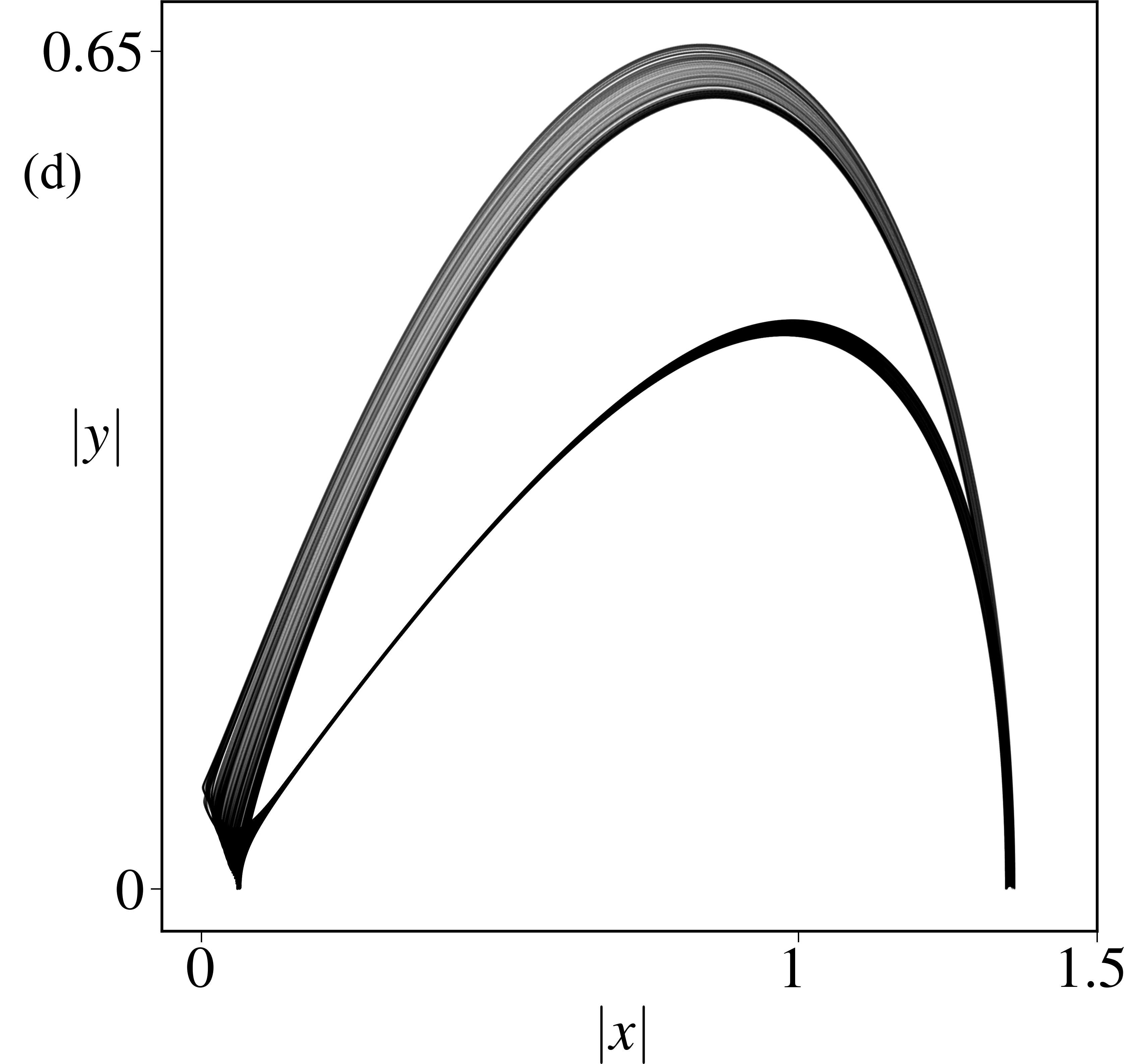}
\caption{
\footnotesize
(a) Projection of the flow uniformly hyperbolic attractor of system~\eqref{eq02} onto the plane of variables $\operatorname{Re}{x}$ and $\operatorname{Im}{x}$.
(b) Projection onto the plane of variables $\operatorname{Re}{x}$ and $\operatorname{Re}{y}$.
(c) Projection onto the plane of variables $\operatorname{Re}{x}$ and $z$.
(d) The dynamics of absolute values $|x|$ and $|y|$.
The parameter values are $\mu=0.98$, $\alpha=0.7$, $\varepsilon=0.1$.
}
\label{fig03}
\end{figure}

Equations~\eqref{eq02} have been solved numerically using Runge -- Kutta 4$^{th}$ order method. 
Fig.~\ref{fig03} shows the attractor of the flow system at $\lambda=0.98$, $\alpha=0.7$, $\varepsilon=0.1$ in different projections.
Fig.~\ref{fig03}(a) demonstrates the projection of the attractor onto the plane of $\operatorname{Re}{x}$ and $\operatorname{Im}{x}$.
One can see that the trajectory runs along straight lines far from the saddle and turns to different directions only near the saddle. 
Fig.~\ref{fig03}(b) shows different projection onto the $\operatorname{Re}{x}$ and $\operatorname{Re}{y}$: the portrait of attractor visually resembles the filled figure-8.
Between successful scatterings on the saddle the trajectory revolves around points of circle of equilibria $\left(\sqrt{\alpha} e^{i\theta},\,0,\,1\right)$, where $\theta\in\left[0,\,2\pi\right)$. 
Every turn is at different angle $\theta$, thus the projection of the attractor is filled with loops. 
Fig.~\ref{fig03}(c) shows another projection onto the $\operatorname{Re}{x}$ and $z$. 
It is compliant with explanation of panels (a) and (b).
While we consider variable $z$ unimportant for scattering, it is required for return of the trajectory to the vicinity of the saddle. 
Fig.~\ref{fig03}(d) shows the dynamics of the absolute values $|x|$ and $|y|$. 
The trajectory goes from and back to the saddle, but is always at finite distance from it (we have checked it numerically for very long simulation times). 
Trajectories of pseudohyperbolic Lorenz attractor come arbitrarily close to the saddle, on the other hand. 

%%We construct an appropriate Poincar\'{e} cross-section surface of the flow~\eqref{eq02}: $S:=|x|^2-1=0$, trajectories go outwards.
%%It is a hypercylindrical surface with 3D generatrix. 
%%We consider variable $z$ less important, variable $x$ has the same absolute value on every return to the cross-section, 
%%so we trace only variables $\operatorname{Re}{y}_n$, $\operatorname{Im}{y}_n$ and $\theta_n=\arg{x}$.

\begin{figure}[!t]
\centering
\includegraphics[width=.218\textwidth,keepaspectratio]{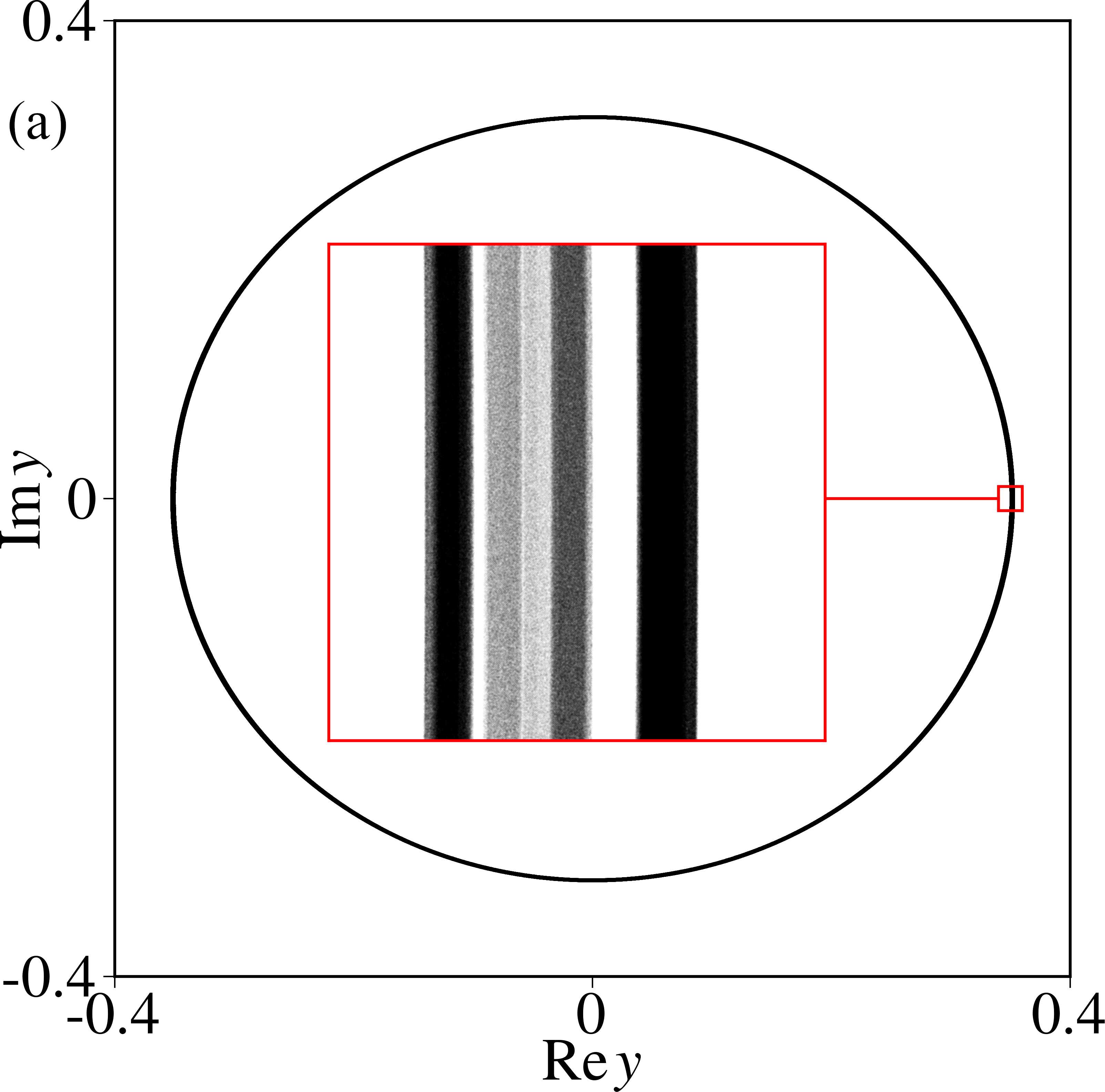}
\includegraphics[width=.215\textwidth,keepaspectratio]{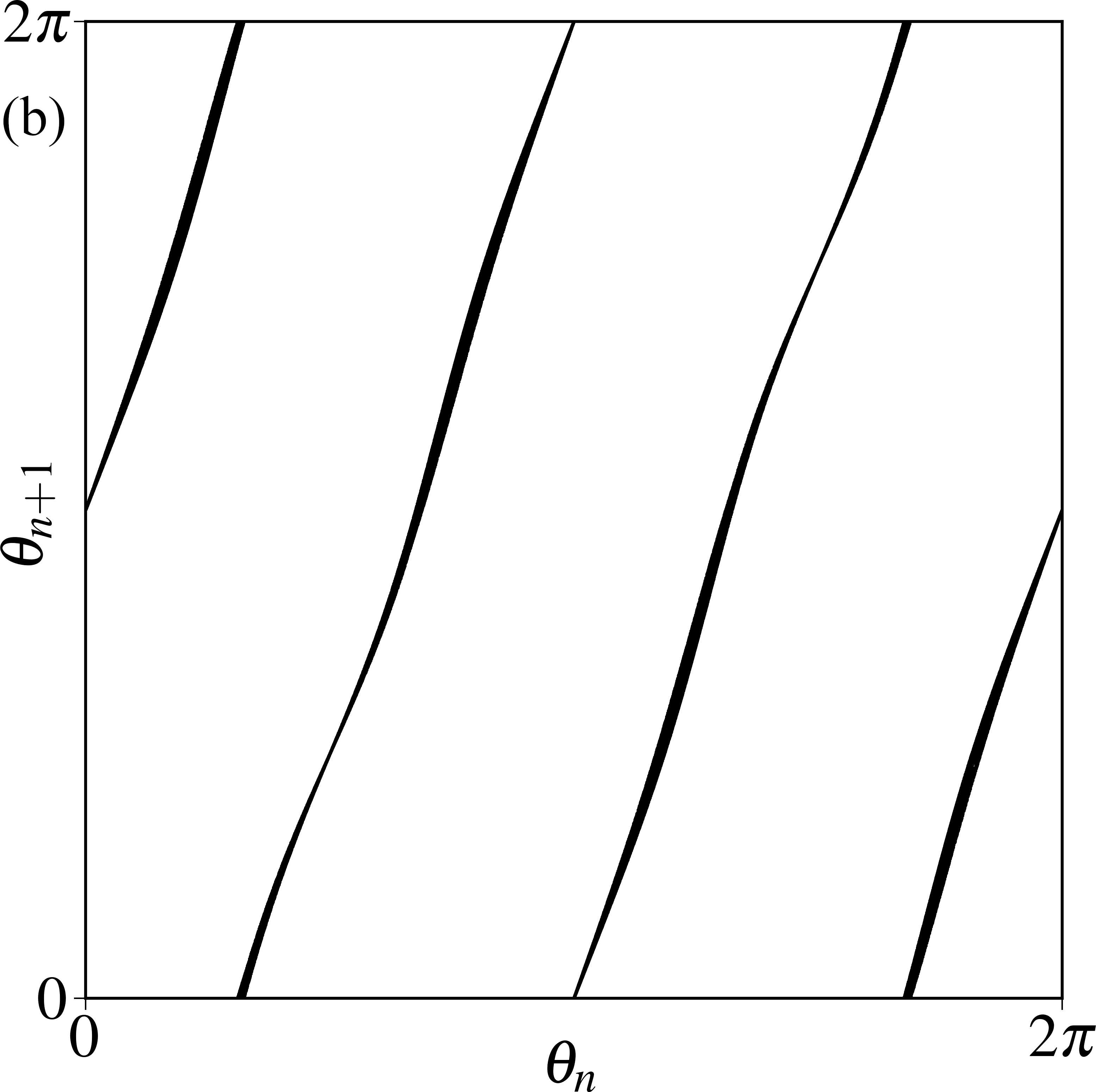}
\includegraphics[width=.22\textwidth,keepaspectratio]{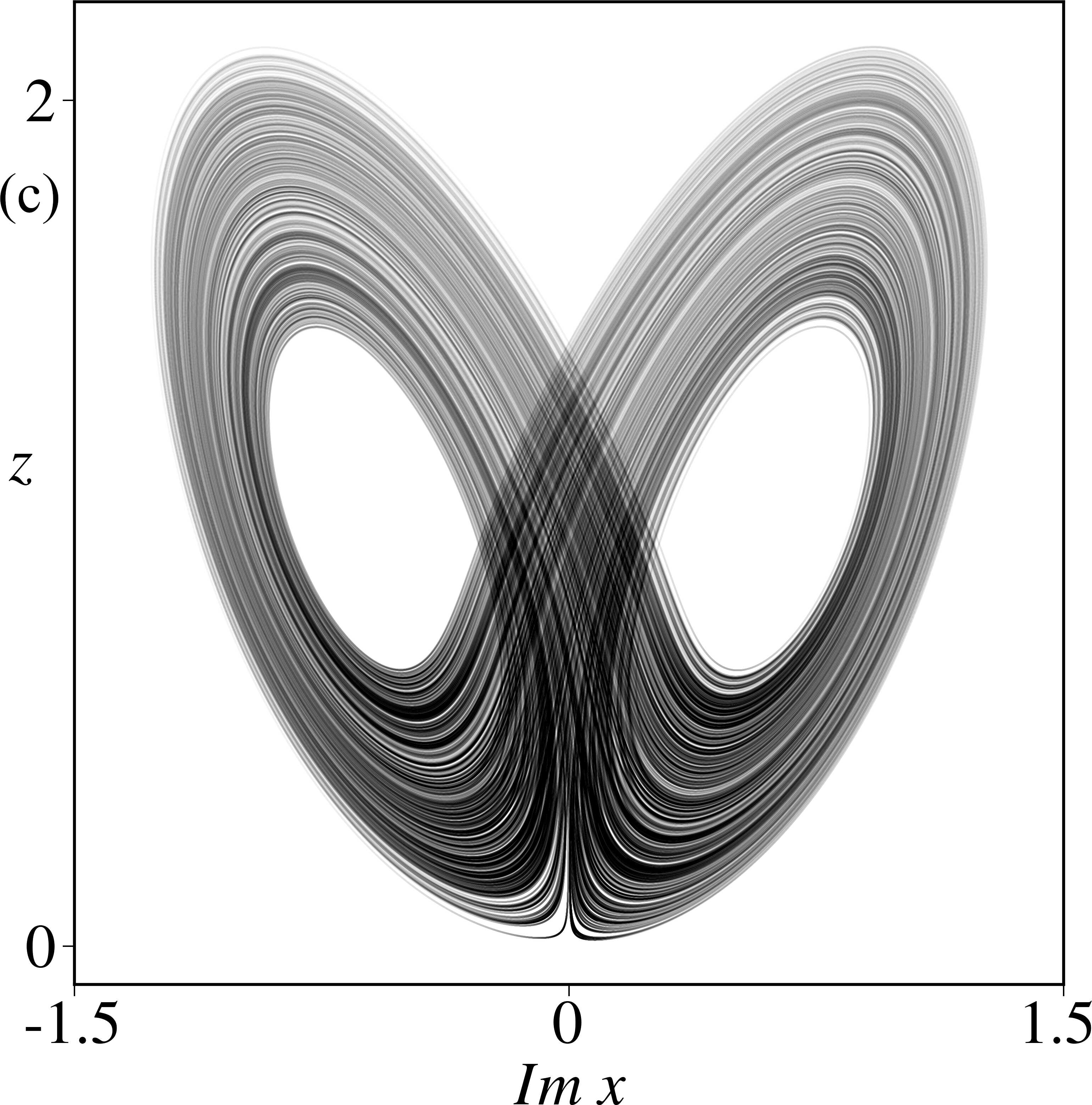}
\includegraphics[width=.24\textwidth,keepaspectratio]{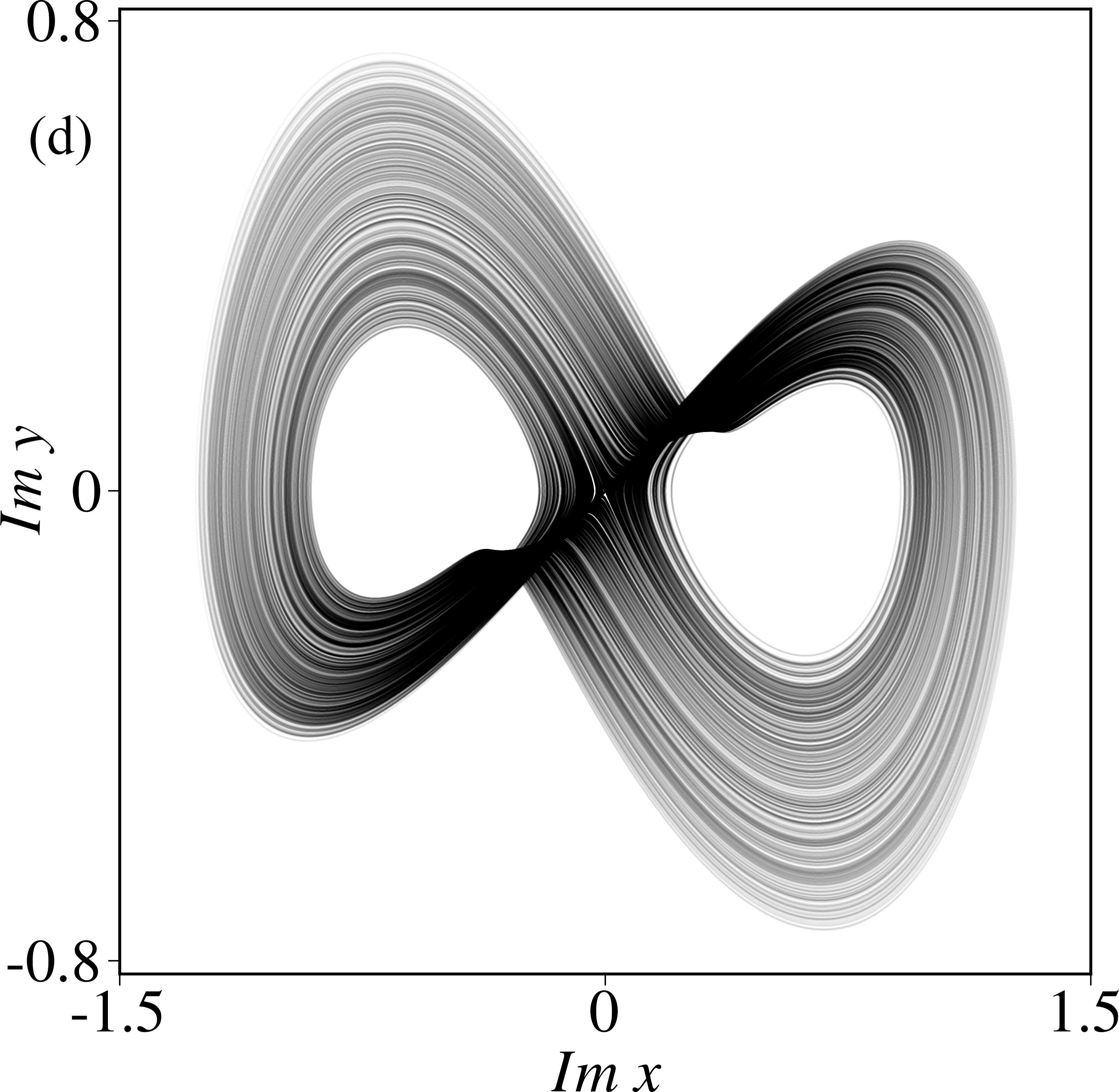}
\caption{
\footnotesize
(a) The portrait of attractor of Poincar\'{e} map for system~\eqref{eq02} and its enlarged part.
(b) The diagram $\theta_{n+1}$ vs. $\theta_{n}$.
The parameter values are $\mu=0.98$, $\alpha=0.7$, $\varepsilon=0.1$.
(c) Projection of the flow non-hyperbolic attractor of system~\eqref{eq02} onto the plane of variables $\operatorname{Im}{x}$ and $z$.
(d) Projection of the attractor of system~\eqref{eq02} onto the plane of variables $\operatorname{Im}{x}$ and $\operatorname{Im}{y}$.
The parameter values are $\mu=0.9$, $\alpha=0.4$, $\varepsilon=0.1$.
}
\label{fig04}
\end{figure}

We construct an appropriate Poincar\'{e} cross-section surface of the flow~\eqref{eq02}: $S:=z-1=0$ ( trajectories go from $z<1$ to $z>1$).
Fig.~\ref{fig04}(a) shows the attractor of Poincar\'{e} return map and its enlarged part for $\lambda=0.98$, $\alpha=0.7$, $\varepsilon=0.1$ (the same values as for Fig.~\ref{fig03}).
We consider it topologically close to Smale -- Williams solenoid on Fig.~\ref{fig01}(b).
The enlarged part demonstrates transversal fractal structure of the attractor.
Fig.~\ref{fig04}(b) shows the diagram $\theta_{n+1}$ vs. $\theta_n$, it is clearly close to Bernoulli map $\theta_{n+1}=3\theta_n+\pi\pmod{2\pi}$.
We have checked numerically that the map for the argument $\theta$ is continuous and monotonous with topological factor of expansion equal to $3$. 
The procedure has been implemented previously~\cite{kuznetsov2021smale}: 
we split the $2\pi$ interval into $N=1000$ pieces, iterate the map $10^6$ times and accumulate the averages $\Phi_k=\frac{1}{T_k}\sum_{n=0}^{T_k} e^{i\theta_n}$
 of $\theta_n$ values, that land into the $k$-th interval, $k\in\left[0,\,N\right]$ is the index of small interval, $T_k$ is the count of $e^{i\theta_n}$ landed into the small interval $k$.
We verify the absence of empty intervals after sufficiently long time of numerical simulation:
if the count $T_k$ of every interval is nonzero, the transformation is continuous.
If all angular shifts $\arg{\Phi_{k+1}}-\arg{\Phi_k}$ between neighboring intervals are positive, then the transformation is monotonous.
Finally we calculate the sum $M=\frac{1}{2\pi}\sum_{k=0}^{N-1}\left(\arg{\Phi_{k+1}}-\arg{\Phi_k}\right)$, which is the expanding factor of the transformation.

The Lyapunov exponents for the flow attractor at $\lambda=0.98$, $\alpha=0.7$, $\varepsilon=0.1$ are
\begin{equation}
%%\nonumber
\begin{aligned}
\lambda_1&=0.0959\pm 0.0005,\\
\lambda_2&=0\pm 0.0001,\\
\lambda_3&=-0.126\pm 0.003,\\
\lambda_4&=-1.060\pm 0.001,\\
\lambda_5&=-1.571\pm 0.003.
\end{aligned}
\label{eq04}
\end{equation}
The first exponent is positive, as it should be for chaotic attractor, the second is zero, as it should be for autonomous system, the other are negative.
Note that every contraction is stronger then expansion: $\lambda_1<|\lambda_3|$.
The Lyapunov dimension of the flow attractor is by Kaplan -- Yorke~\cite{frederickson1983liapunov} formula $D_{KY}=2+\frac{\lambda_1+\lambda_2}{|\lambda_3|}=2.761$, which is rather small for 5D system.

The average time interval between successful Poincar\'{e} cross sections is $T_{av}=11.331$.
The largest Lyapunov exponent for Poincar\'{e} map is $\Lambda_1=T_{av}\lambda_1=1.086\approx\ln{3}$ and the $\ln{3}$ is the Lyapunov exponent for 
Bernoulli map with expansion factor $3$. 
Overall, the attractor of Poincar\'{e} map manifests the features of Smale -- Williams solenoid.

Other kinds of chaotic attractors are also possible at different parameters. 
In classical Shimizu -- Morioka system~\eqref{eq01} at $\lambda=0.9$, $\alpha=0.4$ the Lorenz attractor emerges. 
Very similar attractor appears in modified system~\eqref{eq02} at $\lambda=0.9$, $\alpha=0.4$, $\varepsilon=0.1$, see Fig.~\ref{fig04}(c) and (d).
Interestingly the trajectories of this attractor do not oscillate in directions of $\operatorname{Re}{x}$ and $\operatorname{Re}{y}$, only $\operatorname{Im}{x}$ and $\operatorname{Im}{y}$ and $z$ are non-zero, the attractor is visually confined to three-dimensional space. 
Its Lyapunov exponents are
\begin{equation}
%%\nonumber
\begin{aligned}
\lambda_1&=0.0348\pm 0.0003,\\
\lambda_2&=0\pm 0.0001,\\
\lambda_3&=-0.014\pm 0.001,\\
\lambda_4&=-0.905\pm 0.001,\\
\lambda_5&=-1.3548\pm 0.0002.
\end{aligned}
\label{eq05}
\end{equation}
Note that there are contractions weaker then expansion: $\lambda_1>|\lambda_3|$.
The Lyapunov dimension of the flow attractor is by Kaplan -- Yorke formula $D_{KY}=3+\frac{\lambda_1+\lambda_2+\lambda_3}{|\lambda_4|}=3.023$.
It is bigger, then the Lyapunov dimension of uniformly hyperbolic attractor. 
It is even slightly bigger then $3$, although the attractor appears embedded in 3D space. 
It differs also from Kaplan -- Yorke dimension of classical Lorenz attractor~\eqref{eq01} at $\lambda=0.9$, $\alpha=0.4$: 
$D_{KY}=2.031$.
Though the attractor looks very similar to Lorenz attractor, it is quantitatively different. 

\section{\label{sec4:level1} Numerical test of hyperbolicity}

The pivotal feature of uniformly hyperbolic attractors is transversality of stable $E^s$, neutral $E^n$ and unstable $E^u$ subspaces of tangent space at the every point of the attractor~\cite{lai1993often,anishchenko2000studying}.
We apply fast numerical algorithm~\cite{kuptsov2012fast} based on procedures of covariant Lyapunov vectors computation~\cite{kuptsov2012theory}.
For an autonomous flow governed by $\dot{\textbf{x}}=\textbf{f}\left(\textbf{x}\right)$ in $m$-dimensional phase space
we solve numerically $k\leq m-1$ variational equations 
\begin{equation}
\dot{\textbf{u}}_k=\hat{\textbf{J}}\left(\textbf{x}\right)\cdot\textbf{u}_k,
\label{eq06}
\end{equation}
where $\hat{\textbf{J}}\left(\textbf{x}\right)$ is Jacobi matrix of $\textbf{f}\left(\textbf{x}\right)$, along a typical trajectory on attractor.
Perturbation vectors are orthogonalized and normalized regularly with Gram -- Schmidt procedure. 
Thus we obtain and store fields of $k$ perturbation vectors $\textbf{u}_k\left(t\right)$ along a trajectory. 
We also solve $k$ adjacent variational equations 
\begin{equation}
\dot{\textbf{v}}_k=-\hat{\textbf{J}}^\intercal\left(\textbf{x}\right)\cdot\textbf{v}_k
\label{eq07}
\end{equation}
along the same trajectory backwards in time~\footnote{the ``minus'' sign in equation reduces with negative time step; the trajectory should be stored in computer memory while solving in forward time to prevent divergence due to instability of backward time calculation}, where $\hat{\textbf{J}}^\intercal\left(\textbf{x}\right)$ is transposed Jacobi matrix. 
We obtain and store fields of $k$ vectors $\textbf{v}_k\left(t\right)$, which are orthogonal to some subspace of dimension $m-k$. 
%%It is easy to check, that two orthogonal vectors $\delta\textbf{x}$ and $\delta\textbf{y}$, $\delta\textbf{y}^\intercal\cdot\delta\textbf{x}=0$, stay orthogonal along a trajectory: 
%%$\frac{d}{dt}\left(\delta\textbf{y}^\intercal\cdot\delta\textbf{x}\right)=\frac{d}{dt}\delta\textbf{y}^\intercal\cdot\delta\textbf{x}+
%%\delta\textbf{y}^\intercal\cdot\frac{d}{dt}\delta\textbf{x}=-\delta\textbf{y}^\intercal\cdot\hat{\textbf{J}}\cdot\delta\textbf{x}+
%%\delta\textbf{y}^\intercal\cdot\hat{\textbf{J}}\cdot\delta\textbf{x}=0$.
Note also that if one evaluates $k$ Lyapunov exponents alongside solving for $\textbf{u}_k\left(t\right)$ in forward time and for $\textbf{v}_k\left(t\right)$ in backward time, then the Lyapunov exponents must coincide in pairs. 

We compile fields of matrices $\hat{\textbf{U}}\left(t\right)$ and $\hat{\textbf{V}}\left(t\right)$ with $k$ columns from vectors $\textbf{u}_k\left(t\right)$, $\textbf{v}_k\left(t\right)$ and $m$ rows.
We calculate $k\times k$ matrices $\hat{\textbf{P}}\left(t\right)=\hat{\textbf{V}}^\intercal\left(t\right)\cdot\hat{\textbf{U}}\left(t\right)$, that contain some information about local structure of the attractor.
If matrix $\hat{\textbf{P}}$ is singular at some point $\textbf{x}$, then the $k$-dimensional space, spanned by columns of $\hat{\textbf{U}}$, and $m-k$-dimensional space, orthogonal to columns of $\hat{\textbf{V}}$, are tangent at this point.
The matrix is singular, if its determinant is zero, or equivalently if its minimal singular value $\sigma_k$ is zero. 
%%While calculating $m$ perturbation vectors $\delta\textbf{x}$ and $\delta\textbf{y}$ we actually also obtain upper left matrices $\hat{\textbf{P}}\left[1:1,1:1\right]$, $\hat{\textbf{P}}\left[1:2,1:2\right]$ and so on.
The minimal singular values $\sigma_k$ are related to angles between subspaces $\beta_k=\frac{\pi}{2}-\arccos\sigma_k$. 

The algorithm applies for diffeomorphisms $\textbf{x}_{n+1}=\textbf{F}\left(\textbf{x}_n\right)$ too. 
Then the variational equations are 
\begin{equation}
\textbf{u}_{n+1}=\hat{\textbf{J}}\left(\textbf{x}_n\right)\cdot\textbf{u}_n,
\label{eq08}
\end{equation}
where $\hat{\textbf{J}}\left(\textbf{x}\right)$ is Jacobi matrix of $\textbf{F}\left(\textbf{x}\right)$.
The adjacent variational equations in forward time are 
\begin{equation}
\textbf{v}_{n+1}=\hat{\textbf{J}}^{-\intercal}\left(\textbf{x}_n\right)\cdot\textbf{v}_n,
\label{eq09}
\end{equation}
where $\hat{\textbf{J}}^{-\intercal}\left(\textbf{x}\right)$ is inversed and transposed Jacobi matrix.
In backward time equation~\eqref{eq09} transforms to $\textbf{v}_{n}=\hat{\textbf{J}}^\intercal\left(\textbf{x}_n\right)\cdot\textbf{v}_{n+1}$.
If the diffeomorphism is a numerical Poincar\'{e} return map of the autonomous flow, one has to accurately exclude the perturbation vector corresponding to the neutral subspace in flow system. 
To perform this, one could compute Jacobi matrix for Poincar\'{e} map from one successful cross-section to the next. 
One starts with the full set of $m$ orthogonal perturbation vectors of length $1$ (composing an identity matrix for example) as an initial condition and solves~\eqref{eq06} without Gram -- Schmidt procedure. 
At the next cross-section one has the $m\times m$ Jacobi matrix, that maps perturbation vectors for the flow from cross-surface to itself. 
Then one subtracts the vector orthogonal to cross-section from all the other vectors. 
If the cross-section surface is described by rather simple function ($S:=z-1=0$ for example), then one of the columns in Jacobi matrix is a zero vector. 
One excludes it and the corresponding row and gets the Jacobi matrix for Poincar\'{e} return map. 
If one computes the field of $\hat{\textbf{J}}\left(\textbf{x}\right)$ along the trajectory, one also obtains the inverse matrices, so there is no need to solve adjacent equations~\eqref{eq07} backward in time. 

\begin{figure}[!t]
\centering
\includegraphics[width=.49\textwidth,keepaspectratio]{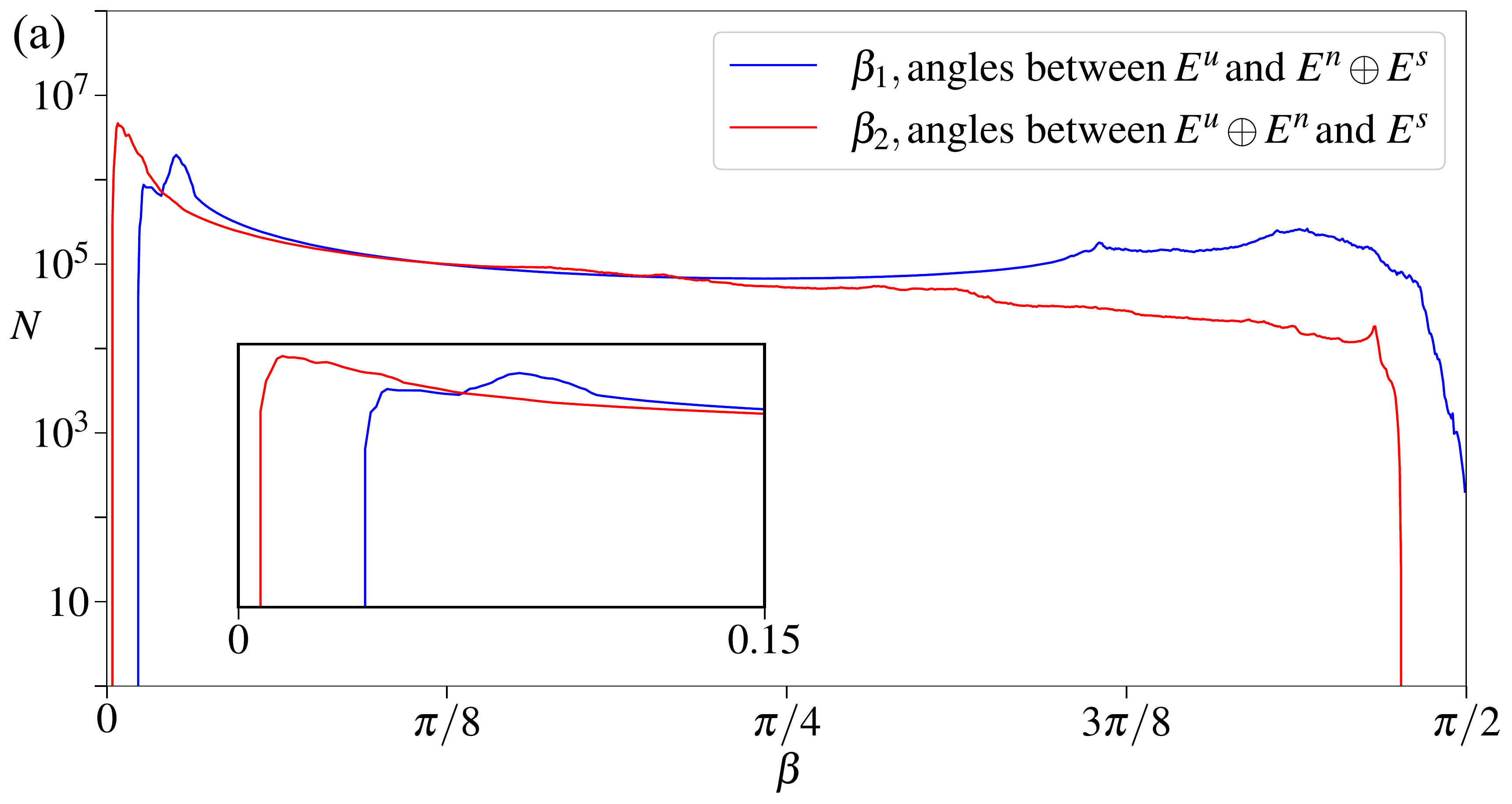}
\includegraphics[width=.49\textwidth,keepaspectratio]{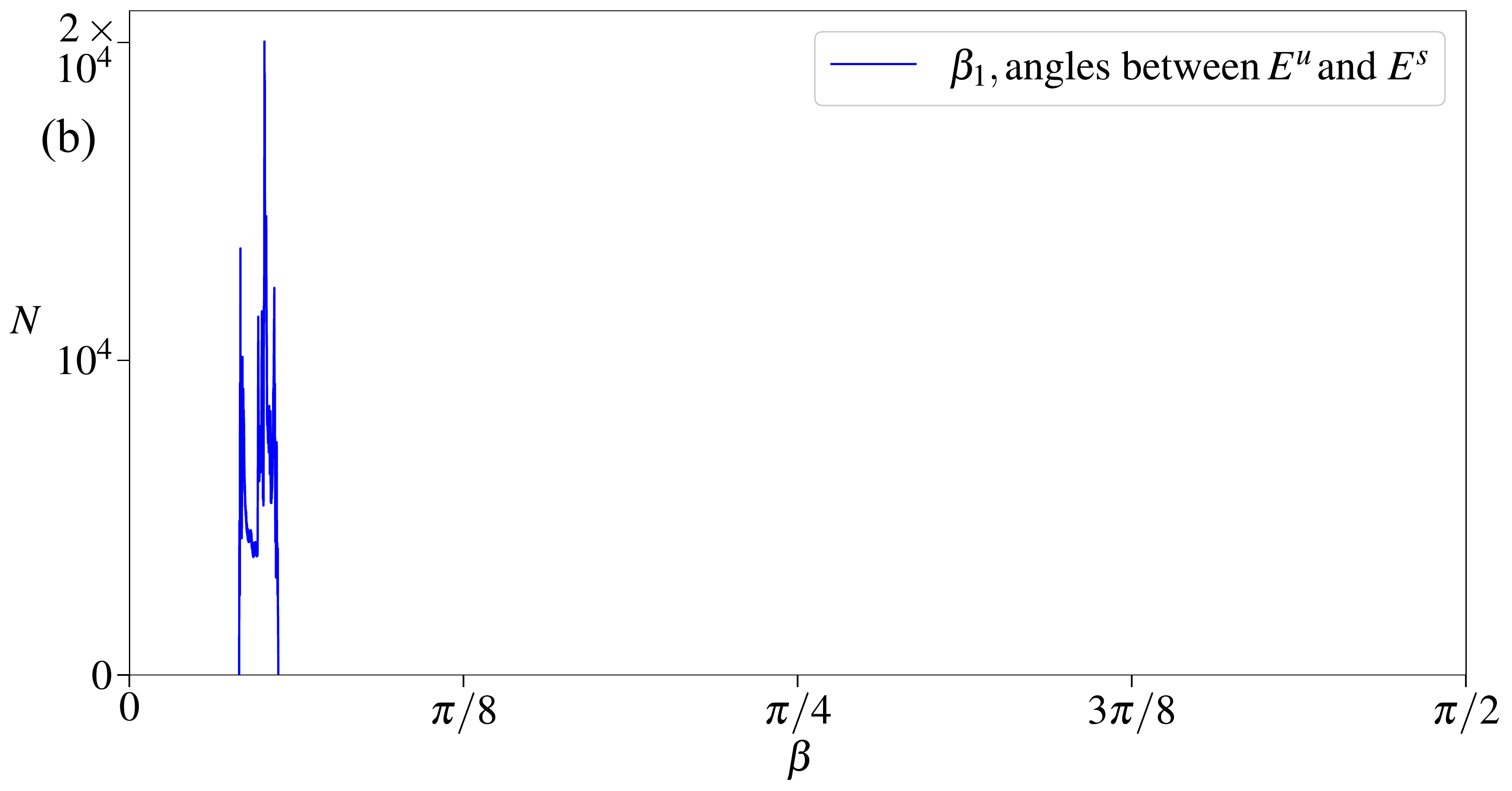}
\caption{
\footnotesize
(a) the distributions of angles between subspaces of the attractor of the flow system~\eqref{eq02} at $\mu=0.98$, $\alpha=0.7$, $\varepsilon=0.1$. 
(b) the distribution of angles between subspaces of the poincar\'{e} map at $\mu=0.98$, $\alpha=0.7$, $\varepsilon=0.1$. 
}
\label{fig05}
\end{figure}

The matrices $\hat{\textbf{P}}\left(t\right)$ converge faster for Poincar\'{e} map, then for the flow, according to our experience. 
However the correct way is to compute the angles between subspaces for the attractor of the flow, because the naive choice of Poincar\'{e} cross-section can lead to incorrect conclusion: some parts of phase space might have tangencies between subspaces and some might not. 

The trajectories of the attractor of the flow~\eqref{eq02} at $\mu=0.98$, $\alpha=0.7$, $\varepsilon=0.1$ have 1-dimensional unstable subspace $E^u$, 
1-dimensional neutral subspace $E^n$ and 3-dimensional stable subspace $E^s$, according to Lyapunov exponents~\eqref{eq04}. 
Therefore it is enough to compute two perturbation vectors in forward and backward time and to obtain $2\times 2$ matrices $\hat{\textbf{P}}\left(t\right)$. 
The upper left elements of $\hat{\textbf{P}}\left(t\right)$ are scalar products of perturbation vectors $\textbf{u}_1\left(t\right)$, spanning the unstable subspace $E^u$, and vectors $\textbf{v}_1\left(t\right)$ orthogonal to the sum of subspaces $E^n\oplus E^s$: $\sigma_1=\textbf{u}_1\left(t\right)\cdot\textbf{v}_1\left(t\right)$. 
Therefore the corresponding angles $\beta_1=\frac{\pi}{2}-\arccos\sigma_1$ are angles between $E^u$ and $E^n\oplus E^s$. 
The smallest singular values $\sigma_2$ of $2\times 2$ matrices $\hat{\textbf{P}}\left(t\right)$ are related to angles $\beta_2=\frac{\pi}{2}-\arccos\sigma_2$ between $E^u\oplus E^n$ and $E^s$: both $\textbf{u}_1$ and $\textbf{u}_2$ span 2-dimensional subspace $E^u\oplus E^n$, $\textbf{v}_1$ and $\textbf{v}_2$ are orthogonal to $E^s$.
We obtain statistics of angles $\beta_1$ and $\beta_2$ for sufficiently long trajectories. If there are zero angles, we conclude that attractor is not hyperbolic. 
If distributions of angles $\beta_1$ and $\beta_2$ are both distanced from zero, then all subspaces are transversal and the attractor of the flow is uniformly hyperbolic according to our numerical approach. 
This also means that the attractor of the Poincar\'{e} cross-section is uniformly hyperbolic too. 
We must be clear, that the absence of zero angles $\beta_k$ in numerical simulations is not a rigorous proof of hyperbolicity. 
Nonetheless this technique lets us distinguish between hyperbolic attractors and quasi-attractors easily and relatively fast.
Rigorous proofs must be based on checking the cone criteria~\cite{shilnikov1997mathematical,wilczak2010uniformly,kuznetsov2007hyperbolic}.

Fig.~\ref{fig05}(a) shows distributions of angles between subspaces of the flow attractor of system~\eqref{eq02} at $\mu=0.98$, $\alpha=0.7$, $\varepsilon=0.1$. 
The blue plot is the distribution of angles $\beta_1$ between unstable subspace $E^u$ and its adjacent subspace $E^n\oplus E^s$. 
The red plot is the distribution of angles $\beta_2$ between subspace $E^u\oplus E^n$ and the stable subspace $E^s$. 
Both distributions are distanced from zero, there is an insert with scaled part of the plots near zero angle. 
The minimal values of angles are $\beta_1^{min}=0.0367$ and $\beta_2^{min}=0.0066$, zero angles are absent
~\footnote{80 trajectories of duration $T_{dur}=10000$ timeunits have been checked to cover all parts of attractor, the timestep of Runge -- Kutta algorithm is $\Delta t=0.001$.}. 
Therefore we conclude that the attractor of the flow system~\eqref{eq02} is uniformly hyperbolic. 
For love of the game we have also calculated the distribution between $E^u$ and $E^s$ subspaces of the attractor of the Poincar\'{e} return map, Fig.~\ref{fig05}(b). 
The minimal value of angle is $\beta_1^{min}=0.1289$ for Poincar\'{e} map
~\footnote{160 trajectories of duration $N_{dur}=10^5$ points have been checked}.

\begin{figure}[!t]
\centering
\includegraphics[width=.45\textwidth,keepaspectratio]{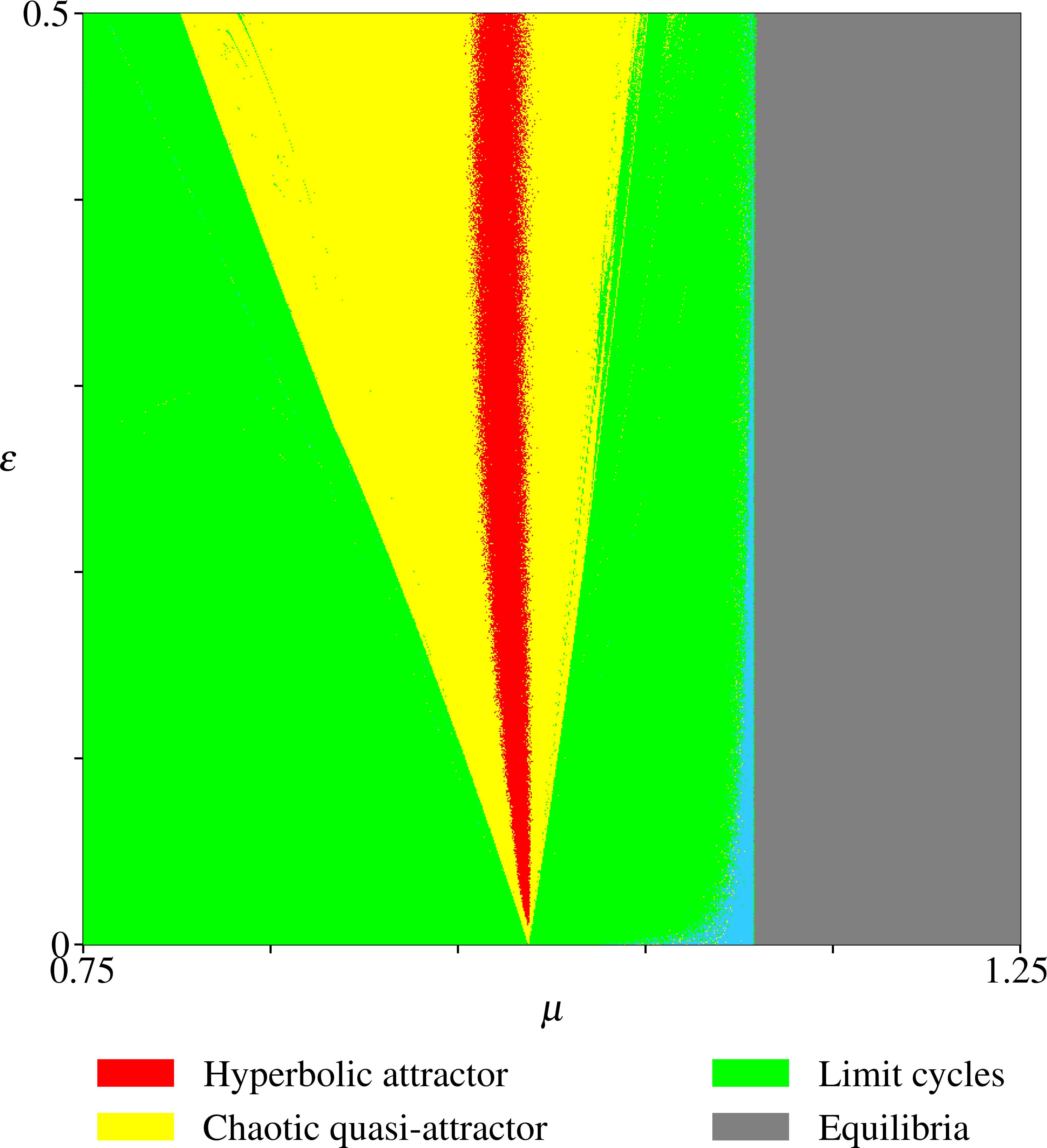}
\caption{
\footnotesize
The chart of regimes of the complex Shimizu -- Morioka model~\eqref{eq02}, $\varepsilon$ vs. $\mu$, obtained numerically by calculating of Lyapunov exponents, by checking of the transversality of tangent subspaces and by checking the expansion factor of the attractor. $\alpha=0.7$.
}
\label{fig06}
\end{figure}

Fig.~\ref{fig06} is a chart of regimes of the flow system~\eqref{eq02}, $\varepsilon$ vs. $\mu$ at fixed $\alpha=0.7$. 
The red region is the domain of uniformly hyperbolic attractors of Smale -- Williams type.
The domain is continuous due to the structural stability of hyperbolic attractor.
Numerous conditions have been checked simultaneously: 
(i) the largest Lyapunov of the flow attractor exponent is positive,
(ii) the angles between subspaces of the flow attractor are never close to zero (the treshold $\beta_{1,2}^{min}>0.001$ is used),
(iii) the expansion factor for the arguments of complex variables of Poincar\'{e} map is $3$. 
The last condition narrows down the domain of hyperbolicity, yet it is required to remove possible non-accurate results of the angle test (the lengths of the checked trajectories are finite, possible zero angles may be missed). 
Smale -- Williams attractors are the only type of hyperbolic attractors that we have found. 
The yellow region is the domain of chaotic quasi-attractors: the angle criteria is violated. 
The green region is the domain of periodic cycles: the largest Lyapunov exponent is zero.
The grey region is the domain of stable equilibria: the largest Lyapunov exponent is negative. 

\begin{figure}[!t]
\centering
\includegraphics[width=.49\textwidth,keepaspectratio]{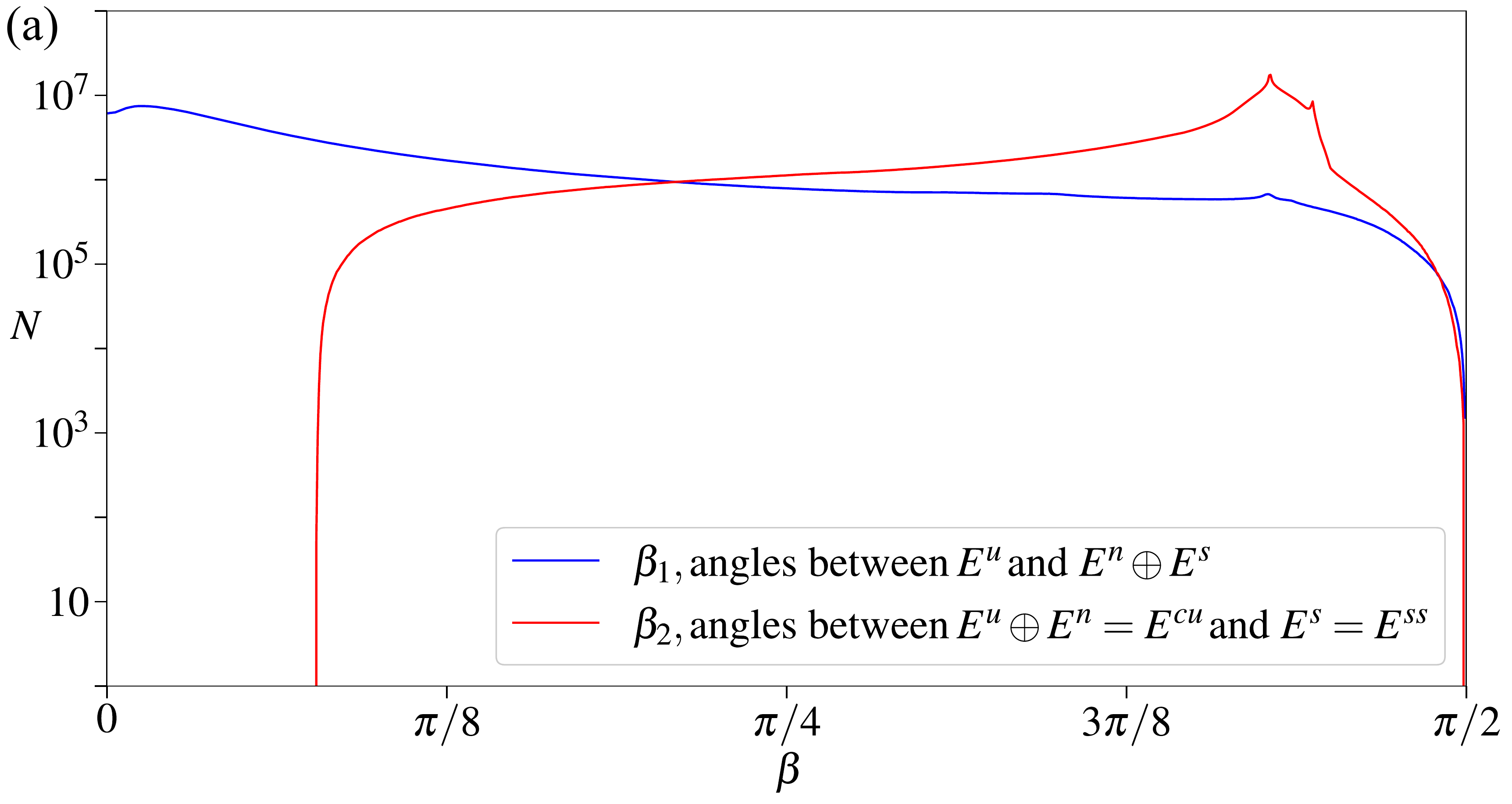}
\includegraphics[width=.49\textwidth,keepaspectratio]{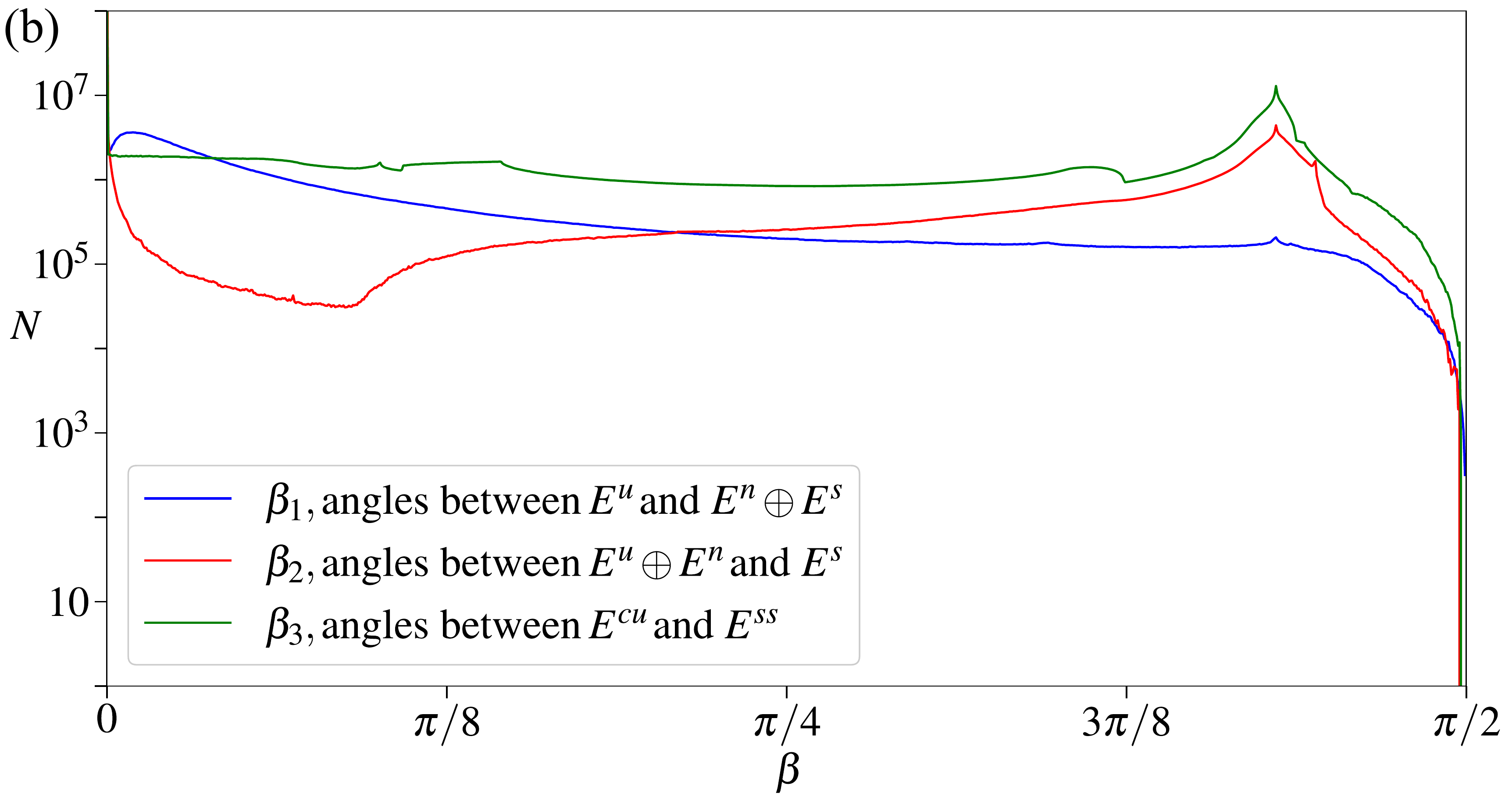}
\caption{
\footnotesize
(a) The distributions of angles between subspaces of the attractor of the classical Shimizu -- Morioka system~\eqref{eq01} at $\mu=0.9$, $\alpha=0.4$. 
(b) The distributions of angles between subspaces of the attractor of the modified complex Shimizu -- Morioka system~\eqref{eq02} at $\mu=0.9$, $\alpha=0.4$. $\varepsilon=0.1$. 
}
\label{fig07}
\end{figure}

The classical Shimizu -- Morioka system~\eqref{eq01} contains genuine pseudohyperbolic Lorenz attractor at $\mu=0.9$, $\alpha=0.4$. 
The conditions of pseudohyperbolicity are:
\begin{enumerate}[label=(\roman*)]
\item 
the largest Lyapunov exponent is positive for every trajectory of attractor,
\item 
the tangent space splits into strongly stable subspace $E^{ss}$, that contains only strongly contracting directions, and centrally unstable subspace $E^{cu}$, that expands volumes, but may include weakly contracting and neutral directions,
\item 
subspaces $E^{ss}$ and $E^{cu}$ are transversal at every point of attractor.
\end{enumerate}
The classical Shimizu -- Morioka attractor at $\mu=0.9$, $\alpha=0.4$ satisfies all of these conditions: 
(i) the largest Lyapunov exponent is $\lambda_1=0.0419\pm 0.0005>0$,
(ii) the sum $\lambda_1+\lambda_2=0.0419>0$ ($\lambda_2=0$), therefore $E^{cu}$ is 2-dimensional and includes expanding direction and neutral direction, $\lambda_3=-1.3418\pm 0.0004<0$, so $E^{ss}$ is 1-dimensional~
\footnote{Lyapunov exponents of the saddle equilibrium also satisfy  these conditions: $\lambda_1=\frac{-\mu+\sqrt{\mu^2+4}}{2}=0.646586$, $\lambda_2=-\alpha=-0.4$, $\lambda_3=\frac{-\mu-\sqrt{\mu^2+4}}{2}=-1.546586$, the saddle value is $\sigma=\lambda_1+\lambda_2=0.246586>0$},
(iii) subspaces $E^{ss}$ and $E^{cu}$ are transversal (Fig.~\ref{fig07}(a)). 
This is the method that had been applied to reveal the domain of pseudohyperbolic Lorenz attractors on the chart from Fig.~\ref{fig02}. 

Lorenz-like attractor at $\mu=0.9$, $\alpha=0.4$, $\varepsilon=0.1$ is an interesting case (see Fig.~\ref{fig04}(c-d)). 
It satisfies only conditions (i) and (ii) of pseudohyperbolicity. 
The strongly stable subspace $E^{ss}$ is 2-dimensional, the centrally unstable subspace is 3-dimensional according to Lyapunov exponents~\eqref{eq05}: 
$\lambda_1+\lambda_2+\lambda_3=0.0208>0$
~\footnote{This is also true for the saddle equilibrium, its Lyapunov exponents are such that 
$\lambda_1+\lambda_2+\lambda_5=-\mu+\sqrt{\mu^2+4}-\alpha=0.893171>0$, $\lambda_{3,4}=-1.546586<0$}. 
Fig.~\ref{fig07}(b) shows distributions of angles $\beta_{1,2,3}$, importantly the angles $\beta_3$ are zero at some points on attractor, so 
the transversality of subspaces $E^{cu}$ and $E^{ss}$ is violated. 
Therefore the flow attractor at $\mu=0.9$, $\alpha=0.4$, $\varepsilon=0.1$ is not pseudohyperbolic. 

\begin{figure}[!t]
\centering
\includegraphics[width=.45\textwidth,keepaspectratio]{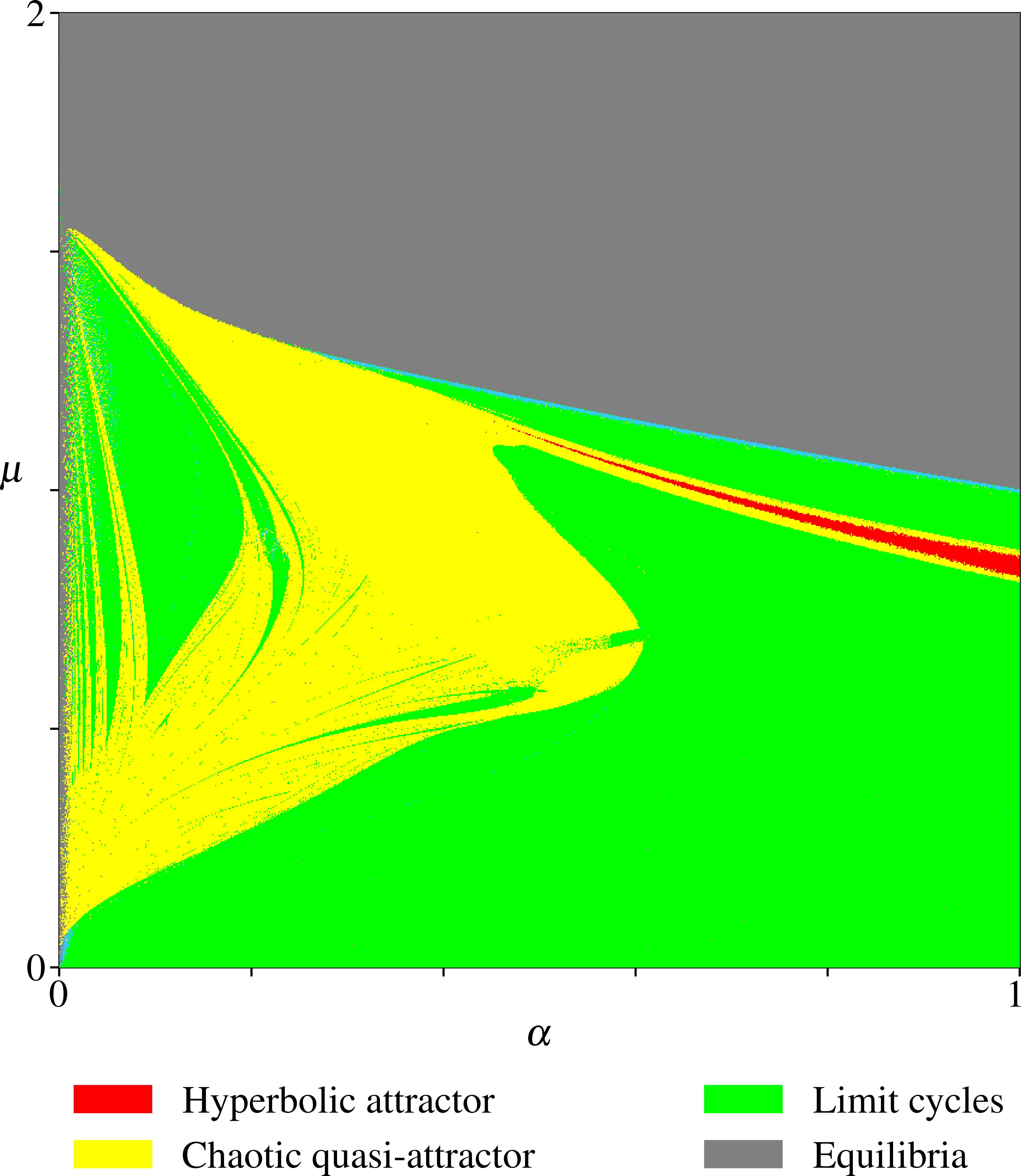}
\caption{
\footnotesize
The chart of regimes of the complex Shimizu -- Morioka model~\eqref{eq02}, $\mu$ vs. $\alpha$, obtained numerically by calculating of Lyapunov exponents, by checking of the transversality of tangent subspaces and by checking the expansion factor of the attractor. $\varepsilon=0.1$. One can compare the chart with Fig.~\ref{fig02}.
}
\label{fig08}
\end{figure}

Fig.~\ref{fig08} is a chart of regimes of the flow system~\eqref{eq02}, $\mu$ vs. $\alpha$ at fixed $\varepsilon=0.1$. 
The red region is the domain of uniformly hyperbolic attractors of Smale -- Williams type.
It is situated near the line of homoclinic bifurcation with negative saddle value in the original Shimizu -- Morioka system~\eqref{eq01}, compare with Fig.~\ref{fig02}. 
Possibility of pseudohyperbolic attractors has been checked as well. 
No pseudohyperbolic attractors have been found in the complex system~\eqref{eq02}. 

\section{\label{sec4:level1} Conclusion}

In this article we have introduced the modified complex Shimizu -- Morioka system with uniformly hyperbolic attractor of Smale -- Williams type. 
Its operation is based on ``scattering'' of trajectories on the saddle equilibrium in complex-valued systems investigated by us before~\cite{kuznetsov2021smale}. 
New example is physically significant, the only additional term not proposed before us is $\varepsilon y^3$. 
We added it not because of physical relevance, but to induce the instability of the angular variable important to construct Smale -- Williams solenoid. 
We surmise that any holomorphic perturbation $h\left(y\right)$ to the second equation of Shimizu -- Morioka system can give rise to Smale -- Williams attractor, although it requires additional investigations. 

We suppose that the investigated mathematical phenomenon is quite general. 
We know of other examples with the same mechanism behind the Smale -- Williams attractor formation~\cite{kuznetsov2007autonomous,kruglov2014attractor,kuznetsov2012hyperbolic}. 

We are working on more rigorous results. 
The provided results are numerical, and the construction and explanations are phenomenological, but we are developing our approach to construct simple piece-wise models, where some parts of phase space with different dynamics are glued. 
This future development is in the spirit of works~\cite{belykh2019lorenz,belykh2021sliding}, where Lorenz attractor is constructed by gluing linear systems of equations, governing different parts of the phase space. 

\begin{acknowledgments}
The work was carried out with the financial support of the Russian Science Foundation, Grant~No.~21-12-00121~(https://www.rscf.ru/project/21-12-00121/).
The authors thank Prof. Pavel Kuptsov for useful discussions.
\end{acknowledgments}

\nocite{*}
\bibliography{shimizumorioka5d.bib}% Produces the bibliography via BibTeX.

\end{document}